\documentclass[12pt]{article}

\usepackage[margin=1in]{geometry}

\usepackage[T1]{fontenc}
\usepackage[utf8]{inputenc}
\usepackage{mathpazo}  
\usepackage{amsmath,amssymb,amsfonts}
\usepackage{longtable,booktabs,array}
\usepackage{multirow}
\usepackage{tabularx}
\usepackage{adjustbox}
\usepackage[ruled,lined,linesnumbered]{algorithm2e}
\SetAlgoNlRelativeSize{0}
\SetAlgoSkip{smallskip}           

\usepackage{graphicx}

\usepackage{xcolor}
\definecolor{nmiblue}{RGB}{0,109,119}
\definecolor{headinggray}{RGB}{80,80,80} 
\definecolor{linkblue}{RGB}{0,109,119}
\usepackage{hyperref}
\usepackage{url}
\usepackage{xurl}
\def\\UrlBreaks{\\do\/\\do-\\do_}
\hypersetup{
  colorlinks=true,
  linkcolor=nmiblue,
  citecolor=nmiblue,
  urlcolor=nmiblue
}

\usepackage[backend=biber,style=nature,maxbibnames=99,doi=true,url=true,isbn=false,sorting=none]{biblatex}
\usepackage[font=small,labelfont=bf,skip=6pt,belowskip=8pt]{caption}

\usepackage{fancyhdr}
\fancypagestyle{plain}{\pagestyle{fancy}}

\usepackage{etoolbox}
\usepackage{titlesec}
\usepackage{needspace}
\titleformat{\section}{\Large\bfseries\color{headinggray}}{\thesection}{0em}{}
\titlespacing*{\section}{0pt}{3ex plus 1ex minus 0.5ex}{1.2ex plus 0.3ex}
\titleformat{\subsection}{\large\bfseries}{\thesubsection}{0em}{}
\titlespacing*{\subsection}{0pt}{2.5ex plus 0.8ex minus 0.3ex}{0.8ex plus 0.2ex}
\titleformat{\subsubsection}{\normalsize\bfseries}{\thesubsubsection}{0em}{}
\titlespacing*{\subsubsection}{0pt}{2ex plus 0.5ex minus 0.2ex}{0.5ex plus 0.1ex}

\usepackage{authblk}
\title{\textbf{Vertical tacit collusion in AI-mediated markets}}
\author{Felipe M. Affonso}
\affil[Spears School of Business, Oklahoma State University, Stillwater, USA]{}
\affil[*Correspondence: phone: +1 (405) 744-1311; email: felipe.affonso@okstate.edu]{}
\date{}

\begin{document}

\maketitle

\begin{abstract}
\setlength{\parskip}{0.5ex}
\noindent AI shopping agents are being deployed to hundreds of millions of consumers, creating a new intermediary between platforms, sellers, and buyers. We identify a novel market failure: vertical tacit collusion, where platforms controlling rankings and sellers controlling product descriptions independently learn to exploit documented AI cognitive biases. Using multi-agent simulation calibrated to empirical measurements of large language model biases, we show that joint exploitation produces consumer harm more than double what would occur if strategies were independent. This super-additive harm arises because platform ranking determines which products occupy bias-triggering positions while seller manipulation determines conversion rates. Unlike horizontal algorithmic collusion, vertical tacit collusion requires no coordination and evades antitrust detection because harm emerges from aligned incentives rather than agreement. Our findings identify an urgent regulatory gap as AI shopping agents reach mainstream adoption.
\end{abstract}

\vspace{1ex}

\section{Introduction}

Delegation pervades economic life, from consumers relying on financial advisors to voters entrusting representatives to principals contracting with agents throughout organizations. When the intermediary to whom decisions are delegated has exploitable biases, parties at different levels of the value chain share a common target, and each benefits from exploitation regardless of what others do. This convergence of incentives can produce coordinated harm without any communication or agreement. We study this phenomenon in AI-mediated consumer markets, where platforms control the display environment through which products are presented and sellers control the product-level inputs that AI agents process.

Major technology firms are deploying AI shopping agents at scale. ChatGPT's shopping features are available to over 700 million weekly users, Amazon reports 250 million customers have used Rufus, and Perplexity enables direct purchases through AI interfaces \cite{openai_introducing_2025,amazon_amazon_2025,perplexity_ai_shopping_2025}. These developments mark a shift in how consumers interact with digital marketplaces. Rather than browsing and comparing products directly, consumers increasingly delegate purchase decisions to AI intermediaries that evaluate options and make or strongly influence final choices by making recommendations.

This delegation creates a novel market structure with three distinct classes of strategic actors. The platform controls the information architecture through which the AI agent encounters products, including ranking algorithms, endorsement badges like ``Amazon's Choice'' or ``Best Seller'' \cite{bairathi_value_2025,lill_product_2024}, and the strategic placement of comparison options. Sellers control the product-level inputs that the AI agent processes, including descriptions optimized for algorithmic interpretation, pricing anchors, and keyword strategies. The AI agent serves as the consumer's delegate, evaluating options through processes that incorporate both legitimate quality-price tradeoffs and systematic biases. The platform economics literature \cite{rochet_platform_2003,armstrong_competition_2006,hagiu_multi-sided_2015} has analyzed two-sided markets where platforms balance buyer and seller interests, but we introduce a key modification: when AI agents mediate consumer decisions, the platform's incentive to serve consumers honestly is attenuated because exploitation of the AI intermediary becomes profitable.

Recent research has documented extensive cognitive-like biases in the large language models that power these shopping agents, and the evidence is both broad and deep. These biases persist across model families, generations, and interface modalities, suggesting they reflect fundamental properties of how these systems process information rather than quirks of particular implementations. Position effects are particularly pronounced and well-documented. Liu et al.\cite{liu_lost_2024} demonstrated that LLMs exhibit strong position biases in retrieval tasks, with accuracy differing by 15-25 percentage points between first and middle positions in context windows. Guo et al.\cite{guo_serial_2024} confirmed these serial position effects across GPT, Llama-2, and T5 model families, while Shi et al.\cite{shi_judging_2025} documented similar patterns in GPT-3.5, GPT-4, Claude-3, and Gemini-Pro. Most directly relevant to our setting, the ACES benchmark evaluated AI shopping agents making actual product choices and found substantial asymmetries: Claude Sonnet 4 selected products from the top half of listings 77\% of the time versus 23\% for the bottom half, with GPT-4.1 and Gemini 2.5 Flash exhibiting similar patterns, and these biases persisted even when visual interfaces were replaced with text-only APIs \cite{allouah_what_2025}. The Magentic Marketplace study found that AI shopping agents exhibit severe first-proposal bias, creating 10-30x advantages for response speed over quality \cite{bansal_magentic_2025}. Beyond position, LLMs exhibit anchoring biases where initial reference points systematically shift subsequent value judgments \cite{binz_using_2023}, susceptibility to framing effects, confirmation bias, and keyword manipulation \cite{echterhoff_cognitive_2024}, and decoy effects consistent with the 20-30\% choice shifts documented in human decision-making \cite{itzhak_instructed_2023,huber_adding_1982,bateman_decoy_2008}. Broader evaluations have found that LLMs exhibit cognitive biases in 40\% of tested scenarios across diverse model families \cite{koo_benchmarking_2024}.

Critically, these biases are far more exploitable than analogous human cognitive biases. Human populations are heterogeneous: different individuals have different susceptibilities, responses depend on context and attention, and strategies that sway some consumers fail on others. However, AI agents exhibit biases uniformly and consistently. The same input reliably produces the same biased output across millions of interactions. When consumers delegate to shared AI systems, this homogeneity concentrates what was a diffuse attack surface into a single predictable target.

This consistency creates systematically exploitable vulnerabilities. Because AI responses are predictable rather than noisy, exploitation requires no prior knowledge of AI psychology or adversarial intent. Profitable strategies emerge through ordinary optimization against observed outcomes. A platform can learn through simple trial and error that weighting seller bids heavily in its ranking algorithm increases revenue. The mechanism is that high-bidding sellers appear in positions that trigger the AI agent's primacy bias, but the platform need not understand this. Similarly, a seller can learn that description optimization with anchoring language and strategic keywords increases win rates, without knowing that the AI agent's manipulation susceptibility makes it favor such listings. Neither party needs to understand the cognitive vulnerabilities they are exploiting. The platform optimizes its ranking system, sellers optimize their listings, and each independently discovers through reinforcement learning that exploiting AI biases is profitable.

When platform and sellers both engage in this exploitation simultaneously, their strategies interact in ways that produce consumer harm exceeding what either could achieve alone. We term this phenomenon ``vertical tacit collusion'' because it involves parties at different levels of the value chain whose independent behaviors converge on joint exploitation of the AI intermediary, without any communication between them (Fig. 1). This mechanism differs from horizontal algorithmic collusion, where competing sellers learn to coordinate prices \cite{ezrachi_virtual_2016,calvano_artificial_2020,klein_autonomous_2021,rocher_adversarial_2023}. In horizontal settings, firms occupy the same market position and control the same strategic instrument, namely price, which means coordination requires solving a collective action problem where each firm must resist the temptation to undercut. In vertical market structures, which characterize most digital marketplaces, platform and sellers occupy structurally different positions and control non-overlapping instruments: the platform sets ranking algorithms while sellers craft product descriptions. Because their incentives are naturally aligned around a common exploitable target, no coordination is required. Each party benefits from exploitation regardless of what the other does, and their joint presence amplifies the harm.
\begin{figure}[htbp]
\centering
\includegraphics[width=0.95\textwidth]{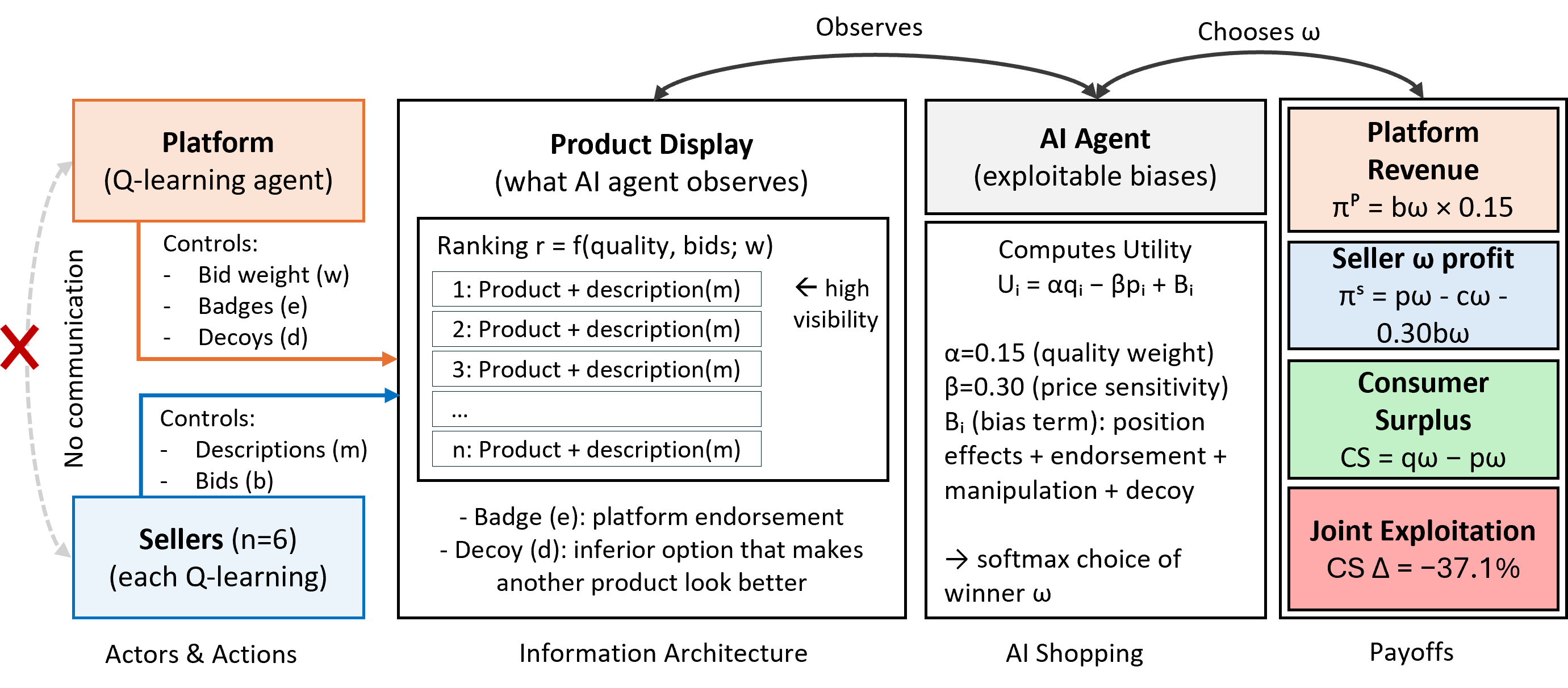}
\caption{\textbf{Vertical tacit collusion in AI-mediated markets.} Conceptual framework showing the three-player game. The platform controls information architecture (ranking algorithm, endorsements, decoys) while sellers control product presentation (descriptions, bids). Both independently learn to exploit the AI shopping agent's cognitive biases without communicating. The AI agent mediates consumer decisions based on perceived utility that incorporates both quality-price tradeoffs and systematic biases. Platform and sellers converge on joint exploitation through independent profit maximization, producing consumer harm that evades traditional antitrust frameworks requiring evidence of coordination.}
\label{fig:framework}
\end{figure}
This alignment creates a regulatory gap with significant implications. 

Antitrust enforcement against collusion typically requires evidence of coordination, whether explicit agreement or tacit understanding through repeated strategic interaction \cite{hovenkamp_federal_2020,whish_competition_2021}. Vertical tacit collusion involves neither. Platform and sellers never communicate, do not observe each other's strategies, and do not condition their behavior on each other's actions. They simply optimize independently against the same exploitable target, and the market structure ensures their strategies align. The consumer harms we document emerge from standard profit-maximizing behavior rather than malicious intent, which means existing competition frameworks designed for horizontal coordination among rivals do not readily address this vertical alignment among non-competitors.

The severity of vertical tacit collusion depends on how platform and seller strategies interact. If both target the same bias channels, their effects could crowd each other out as substitutes, producing joint harm that falls short of the sum of individual effects. This pattern is common in multi-agent reinforcement learning settings \cite{shou_multi-agent_2022,wang_multi-agent_2025}. If each exploits orthogonal vulnerabilities, their effects would simply add. But if platform actions amplify returns to seller manipulation or vice versa, their effects act as complements, producing joint harm that exceeds the sum of parts. We find this third pattern.

We formalize AI-mediated markets as a repeated game \(\Gamma = \langle \mathcal{N}, \mathcal{A}, \mathcal{S}, \pi, \mathcal{T} \rangle\) where \(\mathcal{N} = \{P, S_1, \ldots, S_n, C\}\) denotes the set of players consisting of platform \(P\), sellers \(S_i\), and AI agent \(C\). In each period, the platform selects an information architecture consisting of three components: a bid weight \(w \in \{0, 0.33, 0.67, 1\}\) that determines how rankings blend quality signals with seller payments, an endorsement rule determining which products receive badges, and a decoy decision \(d \in \{0, 1\}\). Sellers simultaneously choose manipulation intensity \(m \in \{0, 1, 2, 3\}\), representing aggressiveness of description optimization through tactics such as keyword insertion, anchoring language, and framing effects \cite{echterhoff_cognitive_2024}, and bid level \(b \in \{0, 1, 2\}\) representing payment for prominence. The AI agent observes the resulting display and computes perceived utility for each product \(i\) as:

\[U_i = \alpha \cdot q_i - \beta \cdot p_i + B_i\]

where \(\alpha\) represents alignment with consumer interests (how much the agent weights quality), \(\beta\) represents price sensitivity, \(q_i\) and \(p_i\) are quality and price, and \(B_i\) aggregates bias effects. The bias term decomposes as:

\begin{align*}
B_i &= \underbrace{\beta_{\text{pos}} \cdot \mathbb{1}(r_i \leq 3) + \beta_{\text{prime}} \cdot \mathbb{1}(r_i = 1) + \beta_{\text{rec}} \cdot \mathbb{1}(r_i = n)}_{\text{position effects}} \\
&\quad + \underbrace{\beta_{\text{end}} \cdot e_i}_{\text{endorsement}} + \underbrace{\beta_{\text{manip}} \cdot m_i \cdot \nu(r_i)}_{\text{manipulation}} + \underbrace{\beta_{\text{decoy}} \cdot d_i}_{\text{decoy}}
\end{align*}

where \(r_i\) is the product's position in the ranking, \(e_i\) indicates endorsement status, \(m_i \in \{0, 1, 2, 3\}\) is manipulation intensity, \(\nu(r_i)\) is position visibility capturing the ``Lost in the Middle'' phenomenon \cite{liu_lost_2024} (manipulation is more effective in high-visibility positions), and \(d_i\) indicates whether a decoy targets product \(i\). This four-channel specification is deliberately parsimonious: each channel corresponds to a well-documented bias class with clear empirical calibration (Supplementary Note 1). We omit additional bias channels (bandwagon effects from stated popularity \cite{koo_benchmarking_2024}, sycophancy toward user framing \cite{sharma_towards_2025}) that would only strengthen our results. The agent selects products according to a softmax distribution:

\[P(\text{choose } i) = \frac{\exp(U_i)}{\sum_{k=1}^{n} \exp(U_k)}\]

\vspace{1ex}

This reduced-form specification captures the essential decision-making properties of LLM shopping agents while enabling systematic analysis across thousands of market interactions that would be computationally prohibitive with actual LLM queries. Parameters are calibrated to match empirical measurements: the ACES benchmark found that real AI shopping agents (Claude, GPT-4.1, Gemini) select top-positioned products at rates implying position effects 8-10 times larger than quality-price variation \cite{allouah_what_2025}, which our parameterization captures (Methods). This logit choice model is standard in discrete choice analysis \cite{zarembka_frontiers_1974} and corresponds to random utility maximization under Type I extreme value errors. Platform and sellers learn through Q-learning following Calvano et al.\cite{calvano_artificial_2020}, a model-free reinforcement learning algorithm that updates action values through temporal difference learning (see Methods for implementation details). Consumer surplus, our primary welfare measure, equals the quality of the chosen product minus its price.

This paper makes three contributions. First, we identify vertical tacit collusion as a distinct anti-competitive phenomenon in AI-mediated markets, demonstrating that platforms and sellers converge on joint exploitation of AI agent biases through independent optimization. Second, we discover a gatekeeper mechanism that determines whether seller strategies help or harm consumers: the platform's ranking choice acts as a binary switch, with seller manipulation benefiting consumers under quality-based ranking but enabling catastrophic harm under bid-based ranking. This mechanism explains the large strategic complementarity we observe and identifies the platform as the critical intervention point. Third, we establish through 82 robustness specifications (including eight learning algorithms, functional form variations, parameter sensitivity tests, and falsification experiments) that vertical tacit collusion reflects genuine economic incentives rather than modeling artifacts, with the effect disappearing only when AI biases are removed. Together, these findings identify a novel form of market failure that current competition policy is not designed to address.

\hypertarget{results}{%
\section{Results}\label{results}}

\phantomsection
\hypertarget{experimental-design}{%
\subsection{Experimental Design}\label{experimental-design}}

We simulate AI-mediated markets consisting of six sellers, one platform,
and one AI shopping agent representing consumers (Methods). Sellers vary
in quality (q $\in$ \{0.90, 0.75, 0.60, 0.45, 0.30, 0.20\}) with
corresponding costs, creating a realistic quality-cost tradeoff
calibrated to gross margins in consumer electronics. The platform and
sellers learn through Q-learning with parameters following Calvano et
al.\cite{calvano_artificial_2020}, while the AI agent follows a fixed policy with biases
calibrated to empirical findings from the LLM cognition literature \cite{liu_lost_2024,binz_using_2023,echterhoff_cognitive_2024,itzhak_instructed_2023}. Each simulation runs for 20,000 rounds, with outcomes
measured over the final 40\% to ensure convergence, and results are
averaged across 100 independent trials with different random seeds.

We compare four experimental conditions to isolate the mechanism of
vertical tacit collusion: a fair-market baseline where the platform
ranks by quality and sellers do not manipulate; platform-only learning
where the platform adapts but sellers remain naive; seller-only learning
where sellers adapt but the platform uses quality-based ranking; and
joint learning where both platform and sellers adapt simultaneously.

\hypertarget{emergence-of-vertical-tacit-collusion}{%
\subsection{Emergence of Vertical Tacit
Collusion}\label{emergence-of-vertical-tacit-collusion}}

Figure 2a presents the central finding (Supplementary Table 2 provides
exact values). In the fair baseline, consumer surplus averages 0.303
(s.d. = 0.002). When both platform and sellers learn to
exploit AI agent biases, consumer surplus falls to 0.191 (s.d. =
0.020), representing a reduction of 37.1\% (P < 0.0001).
This substantial harm emerges without any communication
between platform and sellers, and the effect is observed in 100 of 100
independent trials.

The decomposition in Fig. 2b reveals an asymmetry. Platform-only
learning reduces consumer surplus by 27.0\% relative to baseline (CS =
0.222), as the platform learns to weight bids heavily in rankings.
Seller-only learning, by contrast, increases consumer surplus by 9.6\% (CS
= 0.333). When sellers manipulate without platform cooperation,
manipulation intensity actually correlates with quality \cite{nelson_advertising_1974,kihlstrom_advertising_1984,moorthy_advertising_2000,thomas_empirical_1998}. This occurs because higher-quality sellers earn higher margins, which gives them more resources to invest in description optimization (keyword insertion, anchoring language, persuasive framing). When the platform maintains quality-based ranking, manipulation therefore signals quality rather than substituting for it, and consumers still receive good recommendations. Thus, sellers cannot exploit position bias without platform
cooperation.

This asymmetry produces a result that defies the intuition of additive effects. If platform and seller strategies were independent, joint harm would equal 27.0\% minus 9.6\%, or 17.4\%. Instead, joint harm is 37.1\%, more than double the prediction under independence. The difference of 19.7 percentage points (95\% CI: [18.3, 21.1], $P < 0.0001$, Cohen's $d = 2.75$) represents pure strategic complementarity, harm that exists only because both parties learn simultaneously. The complementarity exceeds the platform's direct effect and substantially exceeds the seller's contribution, reflecting the structural difference between vertical and horizontal multi-agent learning where competing strategies typically crowd each other out. Platform ranking determines which products occupy bias-triggering positions, while seller manipulation determines how effectively those positions convert to sales. This is the signature of vertical tacit collusion: together, platform and sellers produce super-additive harm that neither could generate alone.

\vspace{1.8ex}

\begin{figure}[htbp]
\centering
\includegraphics[width=0.95\textwidth]{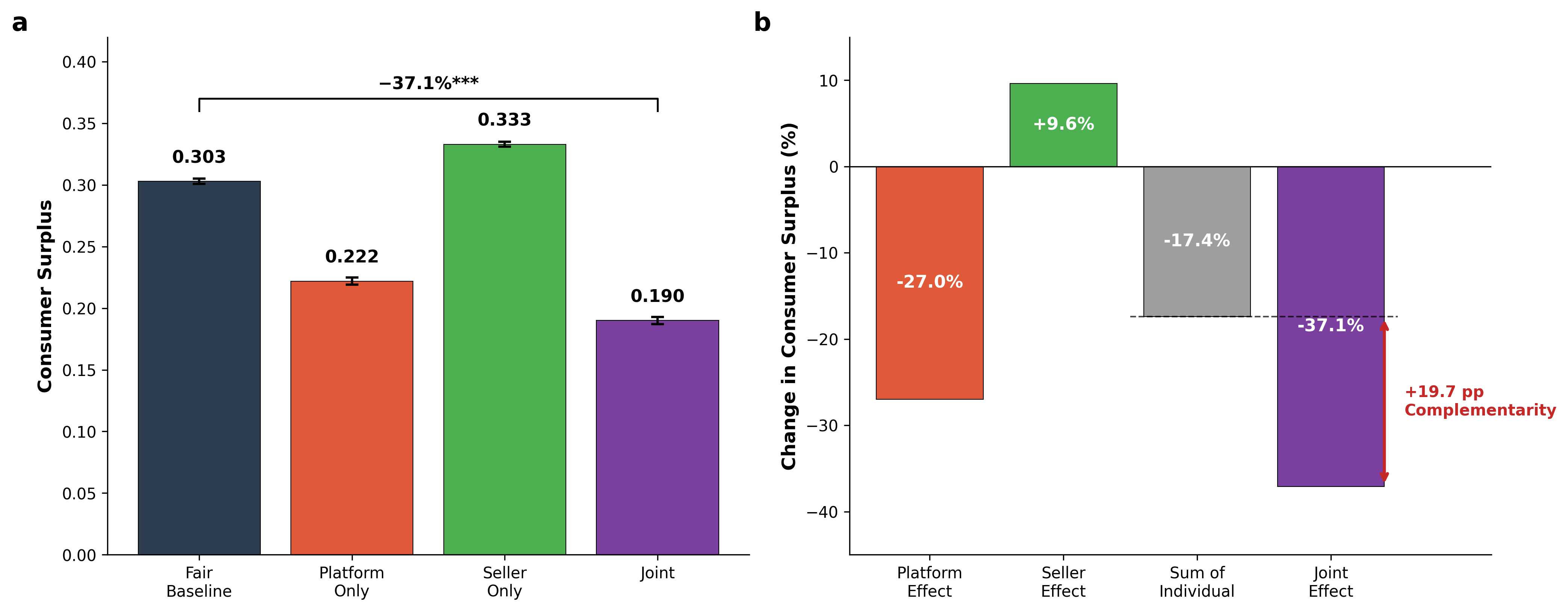}
\caption{\textbf{Emergence of vertical tacit collusion.} \textbf{a}, Consumer surplus across experimental conditions (N = 100 trials). Fair baseline represents an idealized market with quality-based ranking. Joint learning produces 37.1\% reduction relative to baseline ($P < 0.0001$). All 100 trials show positive joint harm. Error bars show 95\% CI. \textbf{b}, Decomposition showing platform-only effect (27.0\% harm), seller-only effect (9.6\% benefit to consumers), and joint effect (37.1\% harm). The gatekeeper mechanism is evident: sellers cannot exploit AI agents without platform cooperation.}
\label{fig:results}
\end{figure}

\hypertarget{the-gatekeeper-mechanism}{%
\subsection{The Gatekeeper
Mechanism}\label{the-gatekeeper-mechanism}}

The seller-only result reveals what we term the
``gatekeeper mechanism.'' Sellers can learn manipulation strategies, but
without platform cooperation through ranking distortion, these
strategies backfire. In the seller-only condition, high-quality sellers
can afford to invest in description optimization while maintaining
competitive prices, and since the platform ranks by quality, consumers
still receive good recommendations. Consistent with the signaling theory of advertising \cite{nelson_advertising_1974,kihlstrom_advertising_1984}, manipulation intensity correlates
positively with quality rather than substituting for it. The quality-win rate correlation under quality-based ranking is 0.76, meaning high-quality sellers win far more often.

When the platform cooperates by weighting bids in rankings, this
relationship inverts. Quality-win correlation drops to 0.56, and the lowest-quality seller's win rate increases from 5.5\% to 18.6\%. Now sellers can bid for position regardless of
quality, and manipulation substitutes for rather than complements
quality. The platform acts as a gatekeeper, determining whether seller
manipulation harms or helps consumers. This gatekeeper role explains the
massive complementarity: sellers need the platform to ``unlock'' their
exploitation potential.

To validate this mechanism, we conducted a threshold analysis varying platform bid weight while allowing sellers to learn (Supplementary Note 20). The results confirm the binary nature of the threshold:
\begin{center}
\begin{longtable}[]{lllll}
\toprule
Bid Weight & Seller Effect & 95\% CI & Harm Rate & Interpretation \\
\midrule
\endhead
\bottomrule
\endlastfoot
0.00 & $-$9.0\% & [$-$9.8, $-$8.1] & 4/100 & Pro-consumer \\
0.33 & $-$8.2\% & [$-$9.1, $-$7.3] & 7/100 & Pro-consumer \\
0.50 & $-$7.1\% & [$-$8.0, $-$6.2] & 10/100 & Pro-consumer \\
0.67 & $-$7.6\% & [$-$8.4, $-$6.8] & 4/100 & Pro-consumer \\
1.00 & +69.2\% & [+68.8, +69.6] & 100/100 & Anti-consumer \\
\end{longtable}
\end{center}\vspace{-2.5em}
\noindent\textbf{Table 1. Gatekeeper threshold analysis.} Seller effect on consumer surplus when platform uses fixed bid weight. Seller manipulation helps consumers at every bid weight from 0 to 0.67; only complete abandonment of quality-based ranking (bid weight = 1.0) enables harmful exploitation.
\vspace{3ex}

Seller manipulation \emph{helps} consumers at every bid weight from 0 to 0.67, with effects ranging from $-$7.1\% to $-$9.0\%. Only when the platform completely abandons quality-based ranking (bid weight = 1.0) do seller strategies flip to massively harmful (+69.2\%). The threshold is effectively binary: quality-based ranking transforms seller manipulation into pro-consumer behavior, while pure bid-based ranking enables catastrophic exploitation. This validates the gatekeeper interpretation: the platform controls whether seller strategies help or harm consumers, with no intermediate regime where modest bid weighting enables modest harm. The platform's ranking choice is the switch that determines exploitation.

\hypertarget{position-bias-as-dominant-channel}{%
\subsection{Position Bias as the Dominant Exploitation Channel}\label{position-bias-as-dominant-channel}}

To identify which bias channels drive exploitation, we conducted a full $2^4$ factorial analysis, running all 16 combinations of the four bias channels (position, endorsement, manipulation, decoy) either active or inactive (Supplementary Note 5; Supplementary Table 4). This design isolates the contribution of each channel and their interactions without imposing assumptions about which channels matter.

The factorial reveals position bias as the dominant exploitation channel, an empirical finding that emerged from the simulation rather than being assumed. Position bias alone produces 29.4\% consumer harm, accounting for 79\% of the full model's 37.1\% effect. What happens without position bias is equally informative: when position bias is disabled but manipulation remains active, consumer harm turns negative, with the manipulation-only specification producing $-$10.3\% harm, meaning consumers actually benefit. Without position bias, seller manipulation cannot be leveraged for exploitation. Instead, sellers fight the platform in ways that inadvertently help consumers. Similarly, badge-only produces modest harm (+22.1\%) and decoy-only produces essentially zero ($-$0.7\%). Thus, position bias is the critical mechanism that transforms individually benign or even pro-consumer biases into exploitation tools.

With position bias active, other channels amplify or moderate harm through complex interactions. Adding endorsement to position increases harm markedly (+38.8\% vs. +29.4\%), but adding manipulation to position decreases harm (+27.8\% vs. +29.4\%), suggesting strategic substitution: when sellers have manipulation as an alternative tool, they moderate bidding for position. The full four-channel model (37.1\%) reflects equilibrium across these competing dynamics. These patterns have clear policy implications: position bias is the primary intervention point, and addressing it alone would eliminate nearly 80\% of consumer harm.

\hypertarget{extensions-and-robustness}{%
\subsection{Extensions and Robustness}\label{extensions-and-robustness}}

A falsification test confirms that exploitable AI biases are necessary for vertical tacit collusion. When we replace the biased AI agent with a debiased agent that evaluates products based solely on quality and price, complementarity collapses from 19.7 pp to 0.7 pp, a 96\% reduction (Supplementary Note 13). A stricter test using true uniform random selection (ignoring all product attributes) produces complementarity indistinguishable from zero ($-$0.2 pp). Without systematic biases to exploit, platform and seller strategies become independent and the super-additive harm disappears. Vertical tacit collusion thus requires AI agents that present exploitable decision-making patterns.

Several extensions speak directly to policy. Human oversight provides limited protection: even when consumers independently override 50\% of AI recommendations, substantial harm persists at +15.1\% with complementarity of 7.3 pp (Supplementary Note 11). Heterogeneous consumer populations remain vulnerable, with significant effects observed even when 75\% of consumers have low bias susceptibility (Supplementary Note 14). Naive sellers using fixed strategies produce higher complementarity than learning sellers, indicating that sophisticated competition actually mitigates exploitation by competing away margins (Supplementary Note 18). Collusion also emerges rapidly, with all trials reaching substantial complementarity within 3,500 rounds on average, suggesting that intervention must be proactive rather than reactive (Supplementary Note 19).

The mechanism operates across market structures ranging from 4 to 36 sellers (Table 2). Complementarity increases with market size, from +7.0 pp in concentrated markets to +27.1 pp in fragmented ones, as more sellers create more opportunities for the platform to arbitrage quality differences through ranking manipulation.

\begin{center}
\begin{longtable}[]{llll}
\toprule
N Sellers & Joint Effect & Complementarity & Cohen's d \\
\midrule
\endhead
\bottomrule
\endlastfoot
4 & +17.3\% & +7.0 pp & 1.66 \\
6 & +37.1\% & +19.7 pp & 2.75 \\
10 & +29.8\% & +17.4 pp & 1.78 \\
18 & +26.1\% & +23.3 pp & 2.73 \\
36 & +20.4\% & +27.1 pp & 4.50 \\
\end{longtable}
\end{center}\vspace{-2.5em}
\noindent\textbf{Table 2. Robustness across market structures.} Complementarity increases with market fragmentation as more sellers create more opportunities for ranking manipulation. All effects significant at P < 0.0001.

\vspace{3ex}

We conducted 82 robustness specifications across 19 analyses (Supplementary Table 3). To address concerns that Q-learning produces artificial collusion through exploration decay \cite{den_boer_artificial_2022,abada_collusion_2024}, we implement Actor-Critic, REINFORCE, and Exp3, algorithms designed to prevent horizontal collusion (Supplementary Note 10). Vertical tacit collusion persists across all three, with effect sizes ranging from $d = 2.92$ to $d = 5.88$. The mechanism also persists across all nine combinations of learning rate and discount factor tested, under both multiplicative and additive functional forms for bias interaction (Supplementary Note 6), and with up to 50\% noise in bias parameter calibration (Supplementary Note 7). Long-run simulations extending to 100,000 rounds show complementarity increasing rather than decaying, reaching 21.8 pp with $d = 6.36$ (Supplementary Note 8). When quality weight is tripled to 0.45, complementarity remains substantial at 10.6 pp ($d = 1.96$), indicating that even strongly quality-aligned AI agents remain vulnerable (Supplementary Note 21).

\hypertarget{discussion}{%
\section{Discussion}\label{discussion}}

We have demonstrated that AI-mediated markets enable a previously unrecognized form of anti-competitive behavior. When platforms and sellers independently optimize their strategies in markets where AI agents mediate consumer decisions, their profit-maximizing behaviors converge on joint exploitation of AI vulnerabilities without any communication between them. This exploitation emerges through learning rather than intent: platforms discover that certain ranking configurations increase revenue, sellers discover that certain description patterns increase win rates, and neither requires knowledge of AI bias literature or adversarial purpose. The result is vertical tacit collusion, a form of coordination that arises not from agreement but from aligned incentives around a shared vulnerability.

The gatekeeper mechanism distinguishes vertical tacit collusion from horizontal algorithmic collusion. In horizontal settings, competing sellers face a coordination problem: each benefits from industry-wide high prices but has individual incentive to undercut. Our vertical setting presents no such coordination problem because platform and sellers are not competitors. They occupy different positions in the value chain with non-overlapping instruments, and their incentives are naturally aligned around a common exploitable target. No reward-punishment scheme is needed to sustain cooperation because there is no temptation to defect.

The welfare implications extend beyond consumer harm to encompass substantial deadweight loss. The platform captures modest rents while destroying considerably more total value, imposing negative externalities on sellers who see profits fall even as they participate in exploitation. This inefficiency arises because platform and seller strategies, while individually profitable, are socially wasteful when combined. Resources flow to manipulation and bid optimization rather than quality improvement, and the information architecture that should help consumers find good products instead systematically misleads them.

These dynamics pose immediate challenges for competition policy. Current antitrust frameworks focus on detecting coordination among competitors, with the US Sherman Act and EU competition law prohibiting agreements in restraint of trade and enforcement requiring evidence of communication or facilitating practices \cite{hovenkamp_federal_2020,whish_competition_2021}. Vertical tacit collusion evades this framework entirely because the platform optimizes its ranking algorithm while sellers optimize their product descriptions, with neither party communicating with the other or even knowing what the other is doing. The harm emerges from aligned incentives rather than coordinated action.

The regulatory challenge is compounded by the legitimate purposes that exploitation strategies serve in isolation. Platforms have valid reasons to incorporate seller payments into ranking decisions since advertising revenue supports platform development, and sellers have valid reasons to optimize product descriptions since compelling presentations can benefit consumers. The problem arises not from any single practice but from how practices interact when deployed against AI agents with exploitable biases. Recent research demonstrates that while LLMs exhibit cognitive biases, these biases often manifest differently than in humans, and models may be irrational in ways that diverge from studied human heuristics \cite{macmillan-scott_irrationality_2024}. This unpredictability compounds the detection challenge because regulators cannot simply apply human-calibrated bias tests to AI agents when the exploitation surfaces may differ from those documented in behavioral economics.

Several policy responses merit consideration. Platforms could face liability for the exploitability of their information architectures, bearing responsibility when their design choices facilitate AI exploitation just as product liability law holds manufacturers responsible for foreseeable misuse. Regulatory bodies could establish minimum robustness standards for commercial AI shopping agents, requiring certification of bias resistance through adversarial evaluation protocols for AI systems that mediate significant consumer transactions \cite{ganguli_red_2022}. The EU AI Act's risk-based classification framework provides a template for such requirements \cite{european_parliament_and_council_of_the_european_union_regulation_2024}. Transparency requirements could mandate disclosure when AI agents encounter information environments that differ from what human consumers would see, analogous to the FTC's endorsement guidelines requiring disclosure of material connections \cite{federal_trade_commission_guides_2023}. Structural remedies could address the alignment of incentives that enables vertical tacit collusion by preventing platforms from simultaneously operating marketplace infrastructure, setting ranking algorithms, and controlling affiliated AI shopping agents.

Beyond competition policy, our findings reveal a vulnerability in current approaches to AI alignment. These approaches focus on ensuring AI systems follow instructions and avoid harmful outputs \cite{ouyang_training_2022,bai_constitutional_2022}, but the three-player framework reveals that AI systems faithfully executing instructions may nonetheless produce harmful outcomes when deployed in adversarial economic environments. An AI shopping agent instructed to find the best product cannot escape manipulation through better instruction-following because the vulnerability lies in how the agent processes strategically optimized inputs. Our sensitivity analysis confirms that improving AI alignment provides limited protection, with consumer harm reducing only modestly as alignment increases while position-based exploitation persists even at high alignment levels. Alignment for economic deployment requires adversarial robustness, specifically the ability to make good decisions regardless of where information appears in context.

Several limitations warrant acknowledgment. Real platforms employ more sophisticated optimization than Q-learning, including A/B testing and deep learning, and real sellers use varied SEO strategies. The mechanism we identify requires only that actors learn from feedback, a feature common to all these approaches. We also model consumers as represented by a single AI agent with fixed biases. In reality, some consumers will continue making their own decisions, and AI agents may improve over time. These dynamics would attenuate harm but would not eliminate it as long as AI agents retain exploitable biases and some consumers delegate decisions to them. Finally, our simulation focuses on the selection stage where AI agents evaluate visible products, while real AI shopping systems also include a retrieval stage where platforms determine which products enter the consideration set \cite{amazon_generative_2024,openai_introducing_2025,google_ai_2025}. Platform gatekeeper power extends beyond ranking to include retrieval itself, suggesting that our findings may be conservative.

The mechanism we identify may operate beyond AI shopping markets. Financial markets with algorithmic trading, healthcare systems with AI diagnostic tools, and legal services with AI document review all involve delegation to intermediaries with potential biases. In each setting, some party controls the information architecture through which the intermediary encounters options. The gatekeeper mechanism suggests this party holds disproportionate power over whether exploitation occurs, making it the natural target for regulatory intervention.

As AI agents assume greater roles in economic life, they will increasingly become targets for strategic manipulation. The vertical structure of digital markets, with platforms controlling information architecture and sellers controlling product presentation, creates natural complementarities in exploitation that emerge without coordination. Addressing this challenge will require regulatory frameworks that look beyond communication-based theories of harm to consider how independent optimization by multiple parties can produce anti-competitive outcomes when those parties share a common, exploitable target.

\hypertarget{methods}{%
\section{Methods}\label{methods}}

\phantomsection
\hypertarget{market-environment}{%
\subsection{Market Environment}\label{market-environment}}

We study a stylized AI-mediated marketplace designed to capture the
essential strategic interactions while remaining analytically tractable.
The market consists of n = 6 sellers, one platform, and one AI shopping
agent representing a population of consumers. This market size reflects
typical consideration set sizes in e-commerce \cite{hauser_evaluation_1990} while enabling
comprehensive robustness analysis across seller counts from 4 to 36.

Our simulation models the selection stage of AI shopping, where the
agent evaluates a consideration set of products and chooses which to
recommend. This is architecturally consistent with deployed systems:
Amazon Rufus retrieves products via retrieval-augmented generation then
reasons over them \cite{amazon_generative_2024}; ChatGPT Shopping searches across
retailers then synthesizes recommendations \cite{openai_introducing_2025}; Google AI
Mode queries its Shopping Graph of 50 billion listings then generates
suggestions \cite{google_ai_2025}. The ACES benchmark \cite{allouah_what_2025}
adopted the same approach, presenting AI agents with 8 products and
measuring choice behavior. We present 6 products, representing a
realistic consideration set size after platform retrieval. The
platform's control over ranking (through bid weight w) captures how
algorithms determine product visibility; sellers' control over
descriptions (through manipulation intensity m) captures how product
listings influence AI interpretation.

Each simulation runs for T = 20,000 rounds. We measure outcomes over the
final 40\% of rounds to ensure convergence, following standard practice
in the algorithmic collusion literature. Results are averaged across 100
independent trials with different random seeds. The relatively simple
structure is deliberate: following Calvano et al.\cite{calvano_artificial_2020}, we seek to
identify whether vertical tacit collusion can emerge in a minimal
setting, establishing the mechanism before introducing complications.

\hypertarget{agents-and-strategies}{%
\subsection{Agents and Strategies}\label{agents-and-strategies}}

Sellers are indexed \(i \in \{1, \ldots, n\}\) with
\(n = 6\). Each seller has quality \(q_i\) and marginal cost \(c_i\):

\begin{center}
\begin{longtable}[]{lllll}
\toprule
Seller \(i\) & Quality \(q_i\) & Cost \(c_i\) & Price \(p_i\) & Margin \\
\midrule
\endhead
\bottomrule
\endlastfoot
1 & 0.90 & 0.15 & 0.49 & 69\% \\
2 & 0.75 & 0.12 & 0.45 & 73\% \\
3 & 0.60 & 0.10 & 0.41 & 76\% \\
4 & 0.45 & 0.08 & 0.38 & 79\% \\
5 & 0.30 & 0.06 & 0.34 & 82\% \\
6 & 0.20 & 0.05 & 0.32 & 84\% \\
\end{longtable}
\end{center}
Higher quality implies higher cost, creating a realistic tradeoff
calibrated to gross margins in consumer electronics \cite{damodaran_margins_2024,ellison_search_2009}.
Prices follow cost-plus pricing:

\[p_i = c_i + \mu_{\text{base}} + \mu_q \cdot q_i\]

with base markup \(\mu_{\text{base}} = 0.25\) and quality premium
\(\mu_q = 0.10\).

The platform selects from \(|\mathcal{A}_P| = 32\)
discrete actions encoding combinations of bid weight
\(w \in \{0, 0.33, 0.67, 1\}\), endorsement rule
\(e \in \{\texttt{quality}, \texttt{bid}, \texttt{hybrid}, \texttt{none}\}\),
and decoy placement \(d \in \{0, 1\}\). The bid weight determines how
rankings blend quality signals with seller payments: at \(w = 0\),
products rank purely by quality; at \(w = 1\), rankings depend entirely
on bids. Each seller selects from
\(|\mathcal{A}_S| = 12\) actions encoding manipulation intensity
\(m \in \{0, 1, 2, 3\}\) and bid level \(b \in \{0, 1, 2\}\):

\begin{center}
\begin{longtable}[]{lll}
\toprule
\(m\) & Description & Components \\
\midrule
\endhead
\bottomrule
\endlastfoot
0 & Honest & No optimization \\
1 & Moderate & AI-text preference \\
2 & Aggressive & + Keywords, anchoring \\
3 & Maximum & + Framing effects \\
\end{longtable}
\end{center}

\hypertarget{ai-agent-decision-model}{%
\subsection{AI Agent Decision Model}\label{ai-agent-decision-model}}

The AI agent computes perceived utility for each product as $U_i = \alpha \cdot q_i - \beta \cdot p_i + B_i$, where $\alpha = 0.15$ represents baseline alignment with consumer interests, $\beta = 0.30$ represents price sensitivity, and $B_i$ aggregates bias effects. The calibrated position effects (primacy + row bonus totaling 1.30) dominate the base quality-price utility range (approximately 0.15), consistent with empirical findings that AI shopping agents select top-positioned products at rates far exceeding rational choice predictions. The ACES benchmark documented 77\% selection rates for top-half positions versus 23\% for bottom-half positions across Claude, GPT-4.1, and Gemini \cite{allouah_what_2025}. In the softmax choice model, this 77:23 ratio implies position effects approximately 8-10 times larger than quality-price variation, which our parameters capture. This parameterization implies that position substantially outweighs quality-price considerations in AI agent decisions, a modeling assumption supported by but also limited to current-generation large language models. Sensitivity analysis confirms robustness: complementarity persists even when quality weight is increased to 0.60 (Supplementary Note 16) or tripled to 0.45 with all biases equalized (Supplementary Note 21).

Bias effects are calibrated to empirical findings from the AI cognition
literature (Supplementary Table 1; Supplementary Note 1 provides
detailed calibration rationale). Position bias follows Liu et al.\cite{liu_lost_2024}, who documented 15-25 percentage point accuracy differences
between first and middle positions, with additional validation from Guo et al.\cite{guo_serial_2024}, Shi et al.\cite{shi_judging_2025}, and the ACES benchmark \cite{allouah_what_2025}. We implement primacy bonus 0.40 for the first position,
row bonus 0.90 for the top three positions, and recency bonus 0.15 for
the last position. These are conservative estimates at the lower bound
of empirically documented ranges. Endorsement effects add 1.2 utility
for badges like ``Top Pick'' while subtracting 0.35 for ``Sponsored''
labels (applied when a seller's bid level equals or exceeds 1.0). Manipulation susceptibility (0.50 per level) follows Echterhoff et al.\cite{echterhoff_cognitive_2024}, who documented LLM susceptibility to description
optimization; under our default multiplicative specification, manipulation
effectiveness is modulated by position visibility with interaction
parameter \(\eta = 0.70\), capturing the ``Lost in the Middle''
phenomenon \cite{liu_lost_2024} while ensuring manipulation retains a floor of $(1-\eta)=30\%$ effectiveness even
in low-visibility positions (Supplementary Note 6 tests an additive alternative with equivalent results). The decoy effect follows the classical
asymmetric dominance literature \cite{huber_adding_1982,bateman_decoy_2008}, with Itzhak et al.\cite{itzhak_instructed_2023} confirming that instruction-tuned LLMs
exhibit this bias; we add 0.4 utility to target products when decoys are
present. Table 3 summarizes the calibration.

\begin{center}
\small
\begin{longtable}[]{llll}
\toprule
Bias Channel & Parameter & Value & Source \\
\midrule
\endhead
\bottomrule
\endlastfoot
Position (primacy) & $\beta_{\text{prime}}$ & 0.40 & Liu et al.\cite{liu_lost_2024} \\
Position (row 1-3) & $\beta_{\text{pos}}$ & 0.90 & ACES \cite{allouah_what_2025} \\
Position (recency) & $\beta_{\text{rec}}$ & 0.15 & Liu et al.\cite{liu_lost_2024} \\
Endorsement & $\beta_{\text{end}}$ & 1.20 & Bairathi et al.\cite{bairathi_value_2025}; Lill et al.\cite{lill_product_2024} \\
Sponsored penalty & $\beta_{\text{spon}}$ & $-$0.35 & Eisend\cite{eisend_meta-analysis_2020} \\
Manipulation & $\beta_{\text{manip}}$ & 0.50/level & Echterhoff et al.\cite{echterhoff_cognitive_2024} \\
Decoy & $\beta_{\text{decoy}}$ & 0.40 & Itzhak et al.\cite{itzhak_instructed_2023} \\
\end{longtable}
\end{center}\vspace{-2.5em}
\noindent\textbf{Table 3. Bias parameter calibration.} All values are conservative estimates at the lower bound of empirically documented ranges. See Supplementary Note 1 for detailed calibration rationale.

\vspace{3ex}

The agent selects products according to a softmax distribution over
utilities, the standard logit choice model in discrete choice analysis
\cite{zarembka_frontiers_1974}. The softmax temperature is set to 1.0 (equivalent to
standard logit); we verified robustness to temperature values 0.5 and
2.0 in sensitivity analysis. This probabilistic choice rule ensures that
higher-utility products are more likely to be selected while allowing
for stochastic variation.

\hypertarget{learning-algorithm}{%
\subsection{Learning Algorithm}\label{learning-algorithm}}

Platform and sellers learn through Q-learning, a model-free
reinforcement learning algorithm that estimates action values through
temporal difference updates. Each agent
\(j \in \{P, S_1, \ldots, S_n\}\) maintains a Q-table
\(Q_j: \mathcal{S} \times \mathcal{A}_j \to \mathbb{R}\) mapping
state-action pairs to expected discounted returns. Supplementary Note 2
provides implementation details; Supplementary Note 4 provides formal
specification.

The state \(s^{(t)} \in \mathcal{S}\) encodes discretized market
conditions by partitioning recent average manipulation into 4 bins and
recent average bids into 4 bins, yielding \(|\mathcal{S}| = 16\) states.
This coarse representation follows Calvano et al.\cite{calvano_artificial_2020} and balances
expressiveness with sample efficiency. Q-values update according to the
temporal difference rule:

\[Q_j(s, a) \leftarrow Q_j(s, a) + \alpha \left[ r_j + \gamma \max_{a'} Q_j(s', a') - Q_j(s, a) \right]\]

with learning rate \(\alpha = 0.12\) and discount factor
\(\gamma = 0.90\). Agents use \(\varepsilon\)-greedy exploration with
\(\varepsilon_t = \max(0.02, 0.25 \times 0.9995^t)\), selecting the
greedy action with probability \(1 - \varepsilon_t\) and exploring
uniformly otherwise.

\hypertarget{experimental-conditions}{%
\subsection{Experimental Conditions}\label{experimental-conditions}}

We compare four experimental conditions to isolate the mechanism of
vertical tacit collusion:

\begin{center}
\begin{tabular}{llll}
\toprule
Condition & Platform & Sellers & Notation \\
\midrule
Fair baseline & \(w = 0\) (quality ranking) & \(m_i = 0, b_i = 0\) &
\(\mathcal{C}_0\) \\
Platform-only & Learning & Naive (\(m_i = 0, b_i = 0\)) &
\(\mathcal{C}_P\) \\
Seller-only & \(w = 0\) (fixed) & Learning & \(\mathcal{C}_S\) \\
Joint learning & Learning & Learning & \(\mathcal{C}_{PS}\) \\
\bottomrule
\end{tabular}
\end{center}

The key test for vertical tacit collusion is whether substantial
consumer harm emerges under joint learning \(\mathcal{C}_{PS}\) without
any communication between platform and sellers. We measure strategic
complementarity as the super-additive component of joint harm: letting
\(\Delta_{CS}(\mathcal{C}) = \overline{CS}_0 - \overline{CS}_\mathcal{C}\)
denote consumer harm under condition \(\mathcal{C}\) relative to
baseline, complementarity is
\(\text{Comp} = \Delta_{CS}(\mathcal{C}_{PS}) - [\Delta_{CS}(\mathcal{C}_P) + \Delta_{CS}(\mathcal{C}_S)]\).
Positive complementarity indicates that joint exploitation exceeds the
sum of individual effects. Supplementary Note 3 provides formal
derivations of welfare measures.

\hypertarget{payoff-structure}{%
\subsection{Payoff Structure}\label{payoff-structure}}

Platform revenue derives from winning bids: the platform receives
commission \(\kappa = 0.50\) on the effective bid cost \(\phi_W = 0.30\)
paid by the winner, consistent with two-sided market pricing structures \cite{rochet_platform_2003,rysman_economics_2009}. Formally, platform profit is
\(\pi_P = b_\omega \cdot \phi_W \cdot \kappa\) where \(b_\omega\) is the
winner's bid level. Seller profit equals sale margin minus bid costs:
winners earn \(\pi_{S_i} = p_i - c_i - b_i \cdot \phi_W\) while losers
pay \(\pi_{S_i} = -b_i \cdot \phi_L\) with \(\phi_L = 0.02\). These
asymmetric bid costs create incentives for strategic bidding: winners
bear meaningful costs while losers face smaller penalties, following the
economic structure of sponsored search auctions \cite{edelman_internet_2007}.

\printbibliography

\hypertarget{competing-interests}{%
\section*{Competing Interests}\label{competing-interests}}

The authors declare no competing interests.

\hypertarget{ethics-statement}{%
\section*{Ethics Statement}\label{ethics-statement}}

This research involves computational simulations only. No human subjects, animals, or personal data were used. The study analyzes strategic behavior in simulated markets and does not involve deception, manipulation of real markets, or any interventions that could affect actual consumers or businesses.

\hypertarget{code-availability}{%
\section*{Code Availability}\label{code-availability}}

All simulation code and analysis scripts are available at \url{https://osf.io/dfgvn/?view_only=2abf097e6c6b40629a5db00decdfc93e}. The repository includes the full simulation environment, all robustness analyses reported in the supplementary materials, and scripts to reproduce all figures and tables.

\section*{Data Availability}
All data are generated by the simulation code; no external datasets were used.

\end{document}


\begin{center}
{\Large\bfseries Supplementary Information for: Vertical tacit collusion in AI-mediated markets}
\end{center}

\vspace{1.5em}

{\large\bfseries Contents}

\vspace{1em}

{\bfseries Supplementary Materials and Methods}

\vspace{0.3em}

\begin{tabular}{@{\hspace{2em}}p{4.5in}@{}r}
Supplementary Note 1: Bias parameter calibration \dotfill & ii \\[0.2em]
Supplementary Note 2: Q-learning implementation \dotfill & iii \\[0.2em]
Supplementary Note 3: Welfare measurement \dotfill & iv \\[0.2em]
Supplementary Note 4: Formal model specification \dotfill & v \\[0.2em]
Supplementary Note 5: Factorial analysis of bias channels \dotfill & vi \\[0.2em]
Supplementary Note 6: Functional form robustness \dotfill & vi \\[0.2em]
Supplementary Note 7: Stochastic biases \dotfill & vii \\[0.2em]
Supplementary Note 8: Long-run dynamics \dotfill & vii \\[0.2em]
Supplementary Note 9: Strategic substitution \dotfill & viii \\[0.2em]
Supplementary Note 10: Algorithm comparison \dotfill & viii \\[0.2em]
Supplementary Note 11: Consumer override \dotfill & ix \\[0.2em]
Supplementary Note 12: Deadweight loss \dotfill & ix \\[0.2em]
Supplementary Note 13: Debiased AI falsification \dotfill & x \\[0.2em]
Supplementary Note 14: Heterogeneous agent populations \dotfill & xi \\[0.2em]
Supplementary Note 15: Position bias magnitude \dotfill & xi \\[0.2em]
Supplementary Note 16: Quality weight sensitivity \dotfill & xii \\[0.2em]
Supplementary Note 17: Platform take rate \dotfill & xii \\[0.2em]
Supplementary Note 18: Naive seller baselines \dotfill & xiii \\[0.2em]
Supplementary Note 19: Convergence speed analysis \dotfill & xiii \\[0.2em]
Supplementary Note 20: Gatekeeper threshold analysis \dotfill & xiv \\[0.2em]
Supplementary Note 21: Equal bias weights analysis \dotfill & xv \\[0.2em]
Supplementary Note 22: Comparison with horizontal algorithmic collusion \dotfill & xvi \\[0.2em]
Supplementary Note 23: Extended review of LLM cognitive biases \dotfill & xvii \\[0.2em]
Supplementary Note 24: Convergence and equilibrium verification \dotfill & xviii \\
\end{tabular}

\vspace{0.8em}

\filbreak
{\bfseries Supplementary Tables}

\vspace{0.3em}

\begin{tabular}{@{\hspace{2em}}p{4.5in}@{}r}
Supplementary Table S1: Complete parameter specification \dotfill & xix \\[0.2em]
Supplementary Table S2: Main results by condition \dotfill & xxi \\[0.2em]
Supplementary Table S3: Comprehensive robustness analyses \dotfill & xxi \\
\end{tabular}

\vspace{1em}

{\bfseries Reading Guide}

\vspace{0.5em}

These supplementary materials are organized to follow the logical structure of the analysis. Notes 1--4 provide foundational materials and methods: bias parameter calibration from the empirical literature, Q-learning implementation details, welfare measurement definitions, and formal model specification. Notes 5--13 present core analyses that test and validate the main mechanism: the factorial design identifying which bias channels drive harm, functional form and parameter robustness tests, algorithm comparisons addressing Q-learning critiques, and the falsification test establishing that AI biases are necessary for the mechanism. Notes 14--21 extend the analysis to policy-relevant variations: heterogeneous populations, parameter sensitivity, platform business models, convergence dynamics, the gatekeeper threshold analysis, and equal bias weight tests. Notes 22--24 provide broader context: comparison with horizontal algorithmic collusion, extended literature review on LLM biases, and equilibrium verification. Tables S1--S3 consolidate parameter specifications, main results, and comprehensive robustness analyses.

For readers with limited time, we suggest the following priority tiers: (1) Essential for evaluating claims: Note 5 (factorial analysis showing position bias accounts for 79\% of harm), Note 10 (algorithm comparison addressing the Q-learning ``mirage'' critique), Note 13 (falsification test with debiased AI), and Note 20 (gatekeeper threshold analysis validating the core mechanism). (2) Robustness verification: Notes 6--9 (functional form, stochastic parameters, long-run stability, boundary conditions) and Note 21 (equal bias weights); Table S3 Panels F, H, K, and R consolidate these results. (3) Policy implications: Notes 11, 14, 17--19 (consumer override, heterogeneous populations, platform take rates, naive sellers, convergence speed). (4) Technical reference: Notes 1--4 for replication, Notes 22--23 for literature context, Note 24 for equilibrium verification.

\newpage

\hypertarget{supplementary-note-1-bias-parameter-calibration}{%
\subsection{Supplementary Note 1: Bias Parameter
Calibration}\label{supplementary-note-1-bias-parameter-calibration}}

We deliberately model only four bias channels (position, endorsement,
manipulation, decoy) despite the empirical literature documenting many
more. This parsimony serves two purposes: tractability and conservative
estimation. Additional documented biases not modeled include bandwagon
effects where stated popularity shifts choices and
salience biases toward longer descriptions \cite{koo_benchmarking_2024}, sycophancy that causes agreement
with user framing \cite{sharma_towards_2025}, and confirmation bias that reinforces prior
statements \cite{echterhoff_cognitive_2024}. Each omitted bias represents an
additional attack surface for strategic exploitation. Our results should
therefore be interpreted as a lower bound on potential harm from
AI-mediated market distortion.

Position bias parameters (primacy = 0.40, row bonus = 0.90, recency =
0.15) are calibrated conservatively at the lower bound of empirical
findings across multiple model families. Liu et al.\cite{liu_lost_2024} documented
15-25 percentage point accuracy differences between first and middle
positions in long contexts, with the ``lost in the middle'' phenomenon
showing both primacy and recency effects. Guo et al.\cite{guo_serial_2024}
confirmed primacy effects across GPT, Llama-2, and T5 model families,
while Shi et al.\cite{shi_judging_2025} found position biases in GPT-3.5, GPT-4,
Claude-3, and Gemini-Pro during pairwise evaluations. Most directly
relevant, the ACES benchmark \cite{allouah_what_2025} documented strong
position biases in Claude Sonnet 4, GPT-4.1, and Gemini 2.5 Flash when
making product choices in e-commerce settings: Claude selected products
from positions 1-4 at 77.3\% versus 22.7\% for positions 5-8, with
biases persisting even in text-only interfaces. Koo et al.\cite{koo_benchmarking_2024} found
that 40\% of comparisons across 16 LLMs exhibited cognitive biases, with
Order Bias prominent especially in models exceeding 40B parameters. Our
conservative position parameters produce bonuses (+1.30 total for first
position) that nonetheless dominate base utility (\textasciitilde0.13
from quality-price), reflecting this robust empirical pattern while
using lower-bound estimates.

The remaining bias parameters are similarly grounded in empirical
research. Manipulation susceptibility (0.50 per level) is calibrated to
Echterhoff et al.\cite{echterhoff_cognitive_2024}, who documented LLM susceptibility to
description optimization, keyword manipulation, and framing effects. We
use a parsimonious linear specification where manipulation bonus scales
directly with manipulation intensity \(m \in \{0, 1, 2, 3\}\).
Endorsement bonus (1.2) reflects field experimental evidence: Bairathi et al.\cite{bairathi_value_2025} document +25\% clicks and +40.6\% orders for endorsed services ($N=598{,}772$); Lill et al.\cite{lill_product_2024} find +15.8 pp click likelihood, with badge removal decreasing click rates by 53\%. Sponsored penalty ($-0.35$) is informed by the Eisend\cite{eisend_meta-analysis_2020} meta-analysis of 61 studies showing disclosure reduces brand attitudes ($r = -0.108$) and credibility ($r = -0.132$); additional support comes from Liu et al.\cite{liu_meta-analysis_2025}. Decoy boost (0.4) is
calibrated to the classical asymmetric dominance literature \cite{huber_adding_1982,bateman_decoy_2008}, which documents 20-30\% choice shifts
toward target alternatives when decoys are introduced; Itzhak et al.\cite{itzhak_instructed_2023} confirmed that instruction-tuned LLMs exhibit this bias, and
MacMillan-Scott and Musolesi\cite{macmillan-scott_irrationality_2024} validated that multiple LLM families
display irrational reasoning consistent with these cognitive biases.
Finally, Q-learning parameters follow Calvano et al.\cite{calvano_artificial_2020}: learning
rate $\alpha$ = 0.12 and discount factor $\gamma$ = 0.90 are within the standard range
($\alpha$ $\in$ {[}0.1, 0.3{]}, $\gamma$ $\in$ {[}0.90, 0.95{]}) used across the algorithmic
collusion literature \cite{calvano_artificial_2020,klein_autonomous_2021}.

\hypertarget{supplementary-note-2-q-learning-implementation}{%
\subsection{Supplementary Note 2: Q-Learning
Implementation}\label{supplementary-note-2-q-learning-implementation}}

We implement a standard Q-learning framework following Calvano et
al.\cite{calvano_artificial_2020}. Q-tables are initialized with small random values drawn
uniformly from \([-0.01, 0.01]\) to break symmetry. We verified
robustness to initialization by testing uniform zero initialization and
larger random ranges (\([-0.1, 0.1]\)), finding results qualitatively
unchanged.

States encode discretized market conditions through a 16-state
representation combining average manipulation (4 bins:
\([0, 0.75), [0.75, 1.5), [1.5, 2.25), [2.25, 3]\)) and average bids (4
bins: \([0, 0.5), [0.5, 1.0), [1.0, 1.5), [1.5, \infty)\)). Recent conditions
are computed as simple averages over the preceding 100 rounds; this
smoothing prevents excessive state switching while remaining responsive
to strategic shifts. During the first 100 rounds before sufficient
history accumulates, state variables default to zero, placing all
agents in the lowest manipulation and bid state.

Exploration follows an \(\varepsilon\)-greedy schedule with
\(\varepsilon_t = \max(0.02, 0.25 \times 0.9995^t)\). The initial
exploration rate of 0.25 ensures broad action coverage, the decay rate
of 0.9995 yields \(\varepsilon \approx 0.02\) by round 15,000, and the
minimum of 0.02 maintains ongoing exploration to prevent lock-in to
suboptimal equilibria.

The main simulation loop proceeds as follows. At each time step \(t\),
the platform and each seller \(i\) observe the current state
\(s^{(t-1)}\) and select actions according to the \(\varepsilon\)-greedy
policy. The platform action \(a_P\) is decoded into bid weight \(w\),
endorsement rule \(e\), and decoy placement \(d\). Each seller action
\(a_{S_i}\) is decoded into manipulation intensity \(m_i\) and bid level
\(b_i\). Rankings are computed according to
\(\text{score}_i = (1 - w) \cdot q_i + w \cdot (b_i / 2)\), and the AI
agent computes utilities and samples a winner \(\omega\) from the
softmax distribution. Payoffs are computed, the state is updated based
on recent market averages, and Q-tables are updated according to the
temporal difference rule. The simulation pseudocode is provided below.

\begin{algorithm}[H]
\caption{Q-learning simulation for vertical tacit collusion}
\SetKwInOut{Input}{Initialize}
\Input{$Q_P[s,a], Q_{S_i}[s,a] \sim \text{Uniform}(-0.01, 0.01)$ for all $s, a$\\
$s^{(0)} \leftarrow (0, 0)$; \quad $\varepsilon \leftarrow 0.25$}
\For{$t = 1$ \KwTo $T$}{
    \tcp{Platform and seller action selection ($\varepsilon$-greedy)}
    $a_P \leftarrow \arg\max_a Q_P[s, a]$ with prob $1-\varepsilon$, else random\;
    \For{each seller $i$}{
        $a_{S_i} \leftarrow \arg\max_a Q_{S_i}[s, a]$ with prob $1-\varepsilon$, else random\;
    }
    \tcp{Decode actions and compute market outcome}
    $(w, e, d) \leftarrow \text{Decode}(a_P)$\;
    $(m_i, b_i) \leftarrow \text{Decode}(a_{S_i})$ for each $i$\;
    Compute rankings, endorsements, utilities\;
    $\omega \leftarrow \text{SoftmaxSample}(\text{utilities})$\;
    \tcp{Compute payoffs and update}
    $r_P \leftarrow b_\omega \cdot \phi_W \cdot \kappa$\;
    $r_{S_i} \leftarrow (p_i - c_i - b_i \cdot \phi_W)$ if $i=\omega$ else $(-b_i \cdot \phi_L)$\;
    $s' \leftarrow \text{Discretize}(\text{avg}(\text{last 100 rounds}))$\;
    \tcp{Q-learning updates}
    $Q_P[s, a_P] \leftarrow Q_P[s, a_P] + \alpha(r_P + \gamma \max_{a'} Q_P[s', a'] - Q_P[s, a_P])$\;
    $Q_{S_i}[s, a_{S_i}] \leftarrow Q_{S_i}[s, a_{S_i}] + \alpha(r_{S_i} + \gamma \max_{a'} Q_{S_i}[s', a'] - Q_{S_i}[s, a_{S_i}])$\;
    $\varepsilon \leftarrow \max(0.02, \varepsilon \times 0.9995)$\;
}
\end{algorithm}

\vspace{1ex}
We measure outcomes over the final 40\% of rounds (rounds 12,000-20,000)
to ensure strategies have stabilized after the exploration phase. By this point,
exploration rate \(\varepsilon\) has decayed to approximately 0.02, and agents
predominantly exploit learned policies.

Simulations run in Python 3.11 using NumPy for vectorized operations.
Each 20,000-round trial completes in approximately 45 seconds on a
single CPU core (Intel i7-12700), with the 100-trial main specification
requiring approximately 75 minutes total. Random seeds are generated
deterministically as \(42 + 100 \times t\) for trial \(t \in \{0, 1, \ldots, 99\}\),
ensuring reproducibility while providing independent randomization across trials.

\hypertarget{supplementary-note-3-welfare-measurement}{%
\subsection{Supplementary Note 3: Welfare
Measurement}\label{supplementary-note-3-welfare-measurement}}

We measure ex-post consumer surplus as realized quality minus realized
price. Formally, in period \(t\), consumer surplus is
\(CS^{(t)} = q_{\omega^{(t)}} - p_{\omega^{(t)}}\) where
\(\omega^{(t)} \in \{1, \ldots, n\}\) is the winning seller chosen by
the AI agent. Average consumer surplus under experimental condition
\(\mathcal{C}\) is computed over the measurement window (final 40\% of
rounds):

\[\overline{CS}_{\mathcal{C}} = \frac{1}{|\mathcal{T}_{\text{measure}}|} \sum_{t \in \mathcal{T}_{\text{measure}}} CS^{(t)}\]

In the fair baseline, the highest-quality seller wins most frequently,
yielding \(\overline{CS} \approx 0.303\). Under joint exploitation,
lower-quality but better-positioned sellers win more often, reducing
surplus to 0.191. Relative consumer harm under condition \(\mathcal{C}\) versus baseline
is
\(\Delta_{CS}(\mathcal{C}) = (\overline{CS}_0 - \overline{CS}_{\mathcal{C}}) / \overline{CS}_0\),
where positive values indicate harm (surplus reduction) and negative
values indicate benefit. Strategic complementarity measures the
super-additive component of joint harm:

\[\text{Comp} = \Delta_{CS}(\mathcal{C}_{PS}) - \left[\Delta_{CS}(\mathcal{C}_P) + \Delta_{CS}(\mathcal{C}_S)\right]\]

\vspace{1ex}

Equivalently, \(\text{Comp} = (\overline{CS}_P + \overline{CS}_S - \overline{CS}_0 - \overline{CS}_{PS}) / \overline{CS}_0\). Conceptually, a transaction exhibits quality inversion if the winning seller's
quality rank exceeds \(n/2\); in a 6-seller market, this occurs when
sellers ranked 4th, 5th, or 6th by quality win. Our primary outcome measure
is consumer surplus, which captures quality inversion effects through its
quality component. Total welfare is the sum
of consumer surplus, platform revenue, and seller profits:
\(W^{(t)} = CS^{(t)} + \pi_P^{(t)} + \sum_{i=1}^{n} \pi_{S_i}^{(t)}\). In our main specification, applying these definitions yields:

\vspace{1.5ex}

\begin{longtable}[]{lll}
\toprule
Condition & \(\overline{CS}\) & Harm \(\Delta_{CS}\) \\
\midrule
\endhead
\bottomrule
\endlastfoot
Baseline (\(\mathcal{C}_0\)) & 0.303 & --- \\
Platform-only (\(\mathcal{C}_P\)) & 0.222 & +27.0\% \\
Seller-only (\(\mathcal{C}_S\)) & 0.333 & -9.6\% \\
Joint (\(\mathcal{C}_{PS}\)) & 0.191 & +37.1\% \\
\end{longtable}

\vspace{1.5ex}

Harm is computed as $(\overline{CS}_0 - \overline{CS}_{\mathcal{C}}) / \overline{CS}_0$; positive values indicate consumer surplus reduction (harm), negative values indicate consumer surplus increase (benefit).

Strategic complementarity is therefore
\(37.1\% - (27.0\% + (-9.6\%)) = 37.1\% - 17.4\% = +19.7\) percentage
points. The 95\% CI \([18.3, 21.1]\) and Cohen's \(d = 2.75\) indicate
that super-additive harm is both large and highly reliable. The key
insight is that sellers acting alone cannot exploit AI agent
biases; they require platform cooperation through ranking distortion.
This gatekeeper mechanism explains why complementarity is so large. The
platform unlocks exploitation potential that sellers cannot access
independently.

\hypertarget{supplementary-note-4-formal-model-specification}{%
\subsection{Supplementary Note 4: Formal Model
Specification}\label{supplementary-note-4-formal-model-specification}}

We provide complete formal definitions for all model components,
following the notation conventions of Rocher et al.\cite{rocher_adversarial_2023}. The state space \(\mathcal{S}\) encodes discretized market conditions
observed at the start of each period. We partition recent average
manipulation into 4 bins and recent average bids into 4 bins, yielding \(|\mathcal{S}| = 16\) states. Manipulation bins are $[0, 0.75)$, $[0.75, 1.5)$, $[1.5, 2.25)$, $[2.25, 3]$; bid bins are $[0, 0.5)$, $[0.5, 1.0)$, $[1.0, 1.5)$, $[1.5, 2]$. Recent conditions are computed
as simple averages over the preceding 100 rounds, providing agents with
a smoothed estimate of market dynamics.

Platform actions \(a_P \in \mathcal{A}_P = \{0, 1, \ldots, 31\}\) encode
three dimensions via integer decomposition. The mapping
\(a_P \mapsto (w, e, d)\) extracts decoy placement as
\(d = a_P \mod 2\), endorsement rule as
\(e = \lfloor(a_P / 2) \mod 4\rfloor\) with \(e \in \{0, 1, 2, 3\}\),
and bid weight as \(w = \mathcal{W}[\lfloor a_P / 8 \rfloor]\) where
\(\mathcal{W} = (0, \tfrac{1}{3}, \tfrac{2}{3}, 1)\).

Similarly, seller actions
\(a_S \in \mathcal{A}_S = \{0, 1, \ldots, 11\}\) encode bid level as
\(b = a_S \mod 3\) with \(b \in \{0, 1, 2\}\) and manipulation as
\(m = \lfloor a_S / 3 \rfloor\) with \(m \in \{0, 1, 2, 3\}\). The ranking function determines product display order. Given platform
bid weight \(w\) and seller bids, we compute scores as
\(\text{score}_i = (1 - w) \cdot q_i + w \cdot (b_i / 2)\) where bids
are normalized to \([0, 1]\) by dividing by the maximum bid level.
Products are displayed in descending score order, inducing ranking
\(\mathbf{r} = (r_1, \ldots, r_n)\).

Position visibility \(\nu: \{1, \ldots, n\} \to [0, 1]\) captures the
``lost in the middle'' phenomenon documented by Liu et al.\cite{liu_lost_2024}. We
set \(\nu(1) = 1.0\), \(\nu(2) = 0.75\), \(\nu(3) = \nu(n) = 0.55\), and
\(\nu(r) = 0.30\) for middle positions. Under the multiplicative
specification (baseline), manipulation effectiveness is modulated by
position visibility:

\[\text{manipulation bonus} = \beta_{\text{manip}} \cdot m_i \cdot \left[(1 - \eta) + \eta \cdot \nu(r_i)\right]\]

where \(\eta = 0.70\) is the visibility interaction parameter. This
ensures that manipulation retains 30\% effectiveness even in
low-visibility positions while being amplified in high-visibility
positions, capturing the intuition that optimized descriptions still
have some effect even when not prominently displayed, but benefit most
from prominent placement. Under the additive specification (robustness
test), manipulation operates independently of position.

Endorsement assignment depends on the platform's endorsement rule: under
\(e = 0\) (quality), the highest-quality seller receives the badge;
under \(e = 1\) (bid), the highest bidder receives it; under \(e = 2\)
(hybrid), the highest bidder among sellers with quality \(\geq 0.5\)
receives it; under \(e = 3\) (none), no badges are assigned. When decoy
placement \(d = 1\), the highest bidder receives the decoy boost:
\(d_i = 1\) if \(d = 1\) and \(i = \arg\max_j b_j\), and \(d_i = 0\)
otherwise.

\hypertarget{supplementary-note-5-factorial-analysis-of-bias-channels}{%
\subsection{Supplementary Note 5: Factorial Analysis of Bias
Channels}\label{supplementary-note-5-factorial-analysis-of-bias-channels}}

To identify which bias channels drive consumer harm and how they interact, we conducted a full $2^4$ factorial analysis. We ran all 16 combinations of four bias channels (position, endorsement/badge, manipulation, decoy) either active or inactive, with 100 independent trials per specification. This design makes no assumptions about which channels matter; the relative importance of each channel and their interactions emerge from the simulation. Note that ``position'' encompasses all position-related biases (primacy bonus for position 1, row bonus for positions 1--3, and recency bonus for last position), which are enabled or disabled together as a single channel.

\newpage

\begin{longtable}[]{lllllll}
\toprule
Pos & Badge & Manip & Decoy & Joint Effect & Comp & Cohen's $d$ \\
\midrule
\endhead
\bottomrule
\endlastfoot
--- & --- & --- & --- & $-$0.5\% & +0.7 pp & 0.25 \\
--- & --- & --- & \checkmark & $-$0.7\% & +0.7 pp & 0.24 \\
--- & --- & \checkmark & --- & $-$10.3\% & +21.4 pp & 2.70 \\
--- & --- & \checkmark & \checkmark & $-$10.7\% & +21.5 pp & 3.23 \\
--- & \checkmark & --- & --- & +22.1\% & +5.5 pp & 1.49 \\
--- & \checkmark & --- & \checkmark & +21.0\% & +4.3 pp & 1.32 \\
--- & \checkmark & \checkmark & --- & +14.0\% & +20.4 pp & 2.24 \\
--- & \checkmark & \checkmark & \checkmark & +13.6\% & +19.9 pp & 2.21 \\
\checkmark & --- & --- & --- & +29.4\% & +11.5 pp & 2.05 \\
\checkmark & --- & --- & \checkmark & +29.0\% & +11.1 pp & 2.09 \\
\checkmark & --- & \checkmark & --- & +27.8\% & +28.2 pp & 3.00 \\
\checkmark & --- & \checkmark & \checkmark & +28.8\% & +29.0 pp & 3.57 \\
\checkmark & \checkmark & --- & --- & +38.8\% & +12.4 pp & 2.72 \\
\checkmark & \checkmark & --- & \checkmark & +38.7\% & +12.1 pp & 2.64 \\
\checkmark & \checkmark & \checkmark & --- & +37.5\% & +20.6 pp & 2.65 \\
\checkmark & \checkmark & \checkmark & \checkmark & +37.1\% & +19.7 pp & 2.75 \\
\end{longtable}

Position bias emerges as the dominant exploitation channel. Position bias alone produces 29.4\% consumer harm, accounting for 79\% of the full model's 37.1\% effect. This dominance was an empirical discovery rather than an assumption; while position bias has strong empirical support in the literature \cite{liu_lost_2024,guo_serial_2024}, we did not know ex ante that it would be the primary channel for strategic exploitation.

The factorial reveals an unexpected pattern in specifications without position bias. When manipulation is active but position bias is disabled, consumer harm turns negative: the manipulation-only specification produces $-$10.3\% harm, meaning consumers actually benefit. Adding decoy to manipulation produces similar results ($-$10.7\% harm), and even badge plus manipulation reduces harm from +22.1\% (badge-only) to +14.0\%. Without position bias, seller manipulation cannot be leveraged for exploitation. Instead, sellers fight the platform in ways that inadvertently benefit consumers because manipulation costs resources, and without position bias to amplify its returns, manipulation becomes a competitive tool rather than an exploitation tool.

With position bias active, other channels interact in distinct ways. Adding badge to position increases harm considerably (from +29.4\% to +38.8\%), a 9.4 percentage point amplification that occurs because platform badges direct AI attention to specific positions, reinforcing position-based exploitation. Adding manipulation to position produces a counterintuitive harm reduction (from +29.4\% to +27.8\%), reflecting substitution: when sellers can manipulate descriptions, they moderate position bidding. Adding decoy to position has essentially no effect ($-$0.4 pp), and decoy-only produces negligible harm ($-$0.7\%), indistinguishable from the null model. The decoy effect, while documented in laboratory settings \cite{itzhak_instructed_2023}, contributes minimally to strategic exploitation in our market setting.

These results have direct policy implications. Position bias is the primary intervention point, and addressing it alone would eliminate approximately 80\% of consumer harm. Other bias channels become benign or even beneficial once the exploitable structure of position-based ranking is removed: manipulation actually helps consumers and badge effects become modest. Interventions targeting manipulation or decoy effects alone would be ineffective because these channels derive their exploitative potential from position bias, so targeting them directly addresses symptoms rather than causes.

The emergence of position bias as the dominant channel was not predicted by our model design. We included four channels based on empirical documentation of each bias in the AI cognition literature. That position bias proved decisive while manipulation proved pro-consumer (absent position bias) emerged from the simulation dynamics and represents a substantive empirical finding about how AI biases interact in adversarial market settings.

\hypertarget{supplementary-note-6-functional-form-robustness}{%
\subsection{Supplementary Note 6: Functional Form
Robustness}\label{supplementary-note-6-functional-form-robustness}}

A key methodological concern is whether complementarity depends on the
multiplicative visibility specification. Under the baseline
specification, manipulation effects are modulated by position visibility
with a floor term, motivated by the ``Lost in the Middle'' phenomenon
\cite{liu_lost_2024}: information in low-visibility positions is not
merely down-weighted but effectively ignored. Formally, the bias term
is:

\begin{align*}
B_i^{\text{mult}} &= \beta_{\text{pos}} \cdot \mathbb{1}(r_i \leq 3) + \beta_{\text{prime}} \cdot \mathbb{1}(r_i = 1) + \beta_{\text{rec}} \cdot \mathbb{1}(r_i = n) \\
&\quad + \beta_{\text{end}} \cdot e_i + \beta_{\text{manip}} \cdot m_i \cdot [(1-\eta) + \eta \cdot \nu(r_i)] + \beta_{\text{dec}} \cdot d_i
\end{align*}

where \(\eta = 0.70\) ensures manipulation retains 30\% effectiveness
even in low-visibility positions. We test an additive alternative where
manipulation and visibility contribute independently:

\begin{align*}
B_i^{\text{add}} &= \beta_{\text{pos}} \cdot \mathbb{1}(r_i \leq 3) + \beta_{\text{prime}} \cdot \mathbb{1}(r_i = 1) + \beta_{\text{rec}} \cdot \mathbb{1}(r_i = n) \\
&\quad + \beta_{\text{end}} \cdot e_i + \beta_{\text{manip}} \cdot m_i + \eta \cdot \nu(r_i) + \beta_{\text{dec}} \cdot d_i
\end{align*}

\begin{longtable}[]{lllll}
\toprule
Model & Joint Effect & Complementarity & Cohen's \(d\) & \(p\)-value \\
\midrule
\endhead
\bottomrule
\endlastfoot
Multiplicative & +37.1\% & +19.7 pp & 2.75 & \textless0.0001 \\
Additive & +43.1\% & +15.3 pp & 2.42 & \textless0.0001 \\
\end{longtable}

Complementarity persists under the additive specification (15.3 pp,
\(d = 2.42\), \(P < 0.0001\)). While slightly smaller than the
multiplicative case, both represent massive effect sizes. This confirms
that super-additive harm emerges from strategic equilibrium
selection (platform and sellers independently discovering that joint
exploitation is profitable) rather than from functional form
assumptions.

The additive specification produces higher total joint harm (+43.1\% vs
+37.1\%) but lower complementarity because manipulation remains
effective even in low-visibility positions, increasing the seller-only
effect and reducing the marginal contribution of joint action.

\hypertarget{supplementary-note-7-parameter-uncertainty-and-stochastic-biases}{%
\subsection{Supplementary Note 7: Parameter Uncertainty and
Stochastic
Biases}\label{supplementary-note-7-parameter-uncertainty-and-stochastic-biases}}

One might worry that our results depend on precise bias
calibration. To test robustness to parameter uncertainty, we introduce
stochastic variation in bias parameters, drawing each bias parameter
from a normal distribution centered on our calibrated value with
standard deviation equal to a fraction of the mean.

\begin{longtable}[]{llll}
\toprule
Noise Level ($\sigma$/$\mu$) & Complementarity & 95\% CI & Cohen's d \\
\midrule
\endhead
\bottomrule
\endlastfoot
0\% (deterministic) & +19.7 pp & {[}18.3, 21.1{]} & 2.75 \\
10\% & +18.2 pp & {[}16.5, 19.9{]} & 2.14 \\
20\% & +18.5 pp & {[}16.7, 20.2{]} & 2.09 \\
30\% & +14.6 pp & {[}12.4, 16.8{]} & 1.29 \\
50\% & +15.9 pp & {[}13.1, 18.8{]} & 1.09 \\
\end{longtable}

Results remain highly significant even with 50\% noise in parameter
calibration. At higher noise levels, effect sizes remain large (d
\textgreater{} 1.0) and complementarity remains positive, confirming
robustness to parameter uncertainty far exceeding any reasonable
calibration error.

\hypertarget{supplementary-note-8-long-run-dynamics-and-equilibrium-stability}{%
\subsection{Supplementary Note 8: Long-Run Dynamics and Equilibrium
Stability}\label{supplementary-note-8-long-run-dynamics-and-equilibrium-stability}}

We extend simulations to 100,000 rounds (5$\times$ the baseline) to assess
equilibrium stability. If the exploitative regime were transient, we
would expect harm to diminish as agents continue learning. Summary statistics from 50 trials over 100,000 rounds are shown below:

\begin{longtable}[]{ll}
\toprule
Metric & Value \\
\midrule
\endhead
\bottomrule
\endlastfoot
Joint effect & +36.7\% \\
Complementarity & +21.8 pp \\
Cohen's d & 6.36 \\
Positive trials & 50/50 (100\%) \\
Converged bid weight & 0.73 \\
Converged manipulation & 2.4 \\
\end{longtable}

Complementarity actually increases in the long run (21.8 pp vs 19.7 pp
at 20,000 rounds), with the effect size growing substantially (d = 6.36
vs d = 2.75). All 50 trials show positive complementarity, and the
equilibrium strategies are tightly clustered, indicating that vertical tacit collusion represents a stable equilibrium rather than a
transient phenomenon. The converged strategies show platform bid weights around 0.73 (heavy
reliance on bids rather than quality in rankings), seller manipulation
around 60-65\% of maximum intensity, and modest but consistent bidding.
These strategies persist throughout the extended simulation, with no
sign of unraveling or decay.

\hypertarget{supplementary-note-9-strategic-substitution-at-extreme-bias-levels}{%
\subsection{Supplementary Note 9: Strategic Substitution at Extreme
Bias
Levels}\label{supplementary-note-9-strategic-substitution-at-extreme-bias-levels}}

A notable finding emerges from the bias magnitude analysis. We vary
all bias parameters simultaneously by a common scale factor (i.e., a
multiplier of 0.5$\times$ scales position bias, badge effects, manipulation
susceptibility, and decoy effects each to 50\% of their baseline values).
At moderate bias levels (0.5$\times$-1.0$\times$ baseline), platform and seller strategies are
strong complements, with complementarity ranging from +16.0 pp to +19.7
pp.~However, at extreme bias levels (\textgreater1.25$\times$ baseline),
complementarity becomes negative, indicating strategic substitution.

\begin{longtable}[]{lll}
\toprule
Scale Factor & Complementarity & Interpretation \\
\midrule
\endhead
\bottomrule
\endlastfoot
0.50$\times$ & +16.0 pp & Strong complementarity \\
0.75$\times$ & +19.5 pp & Near-maximum complementarity \\
1.00$\times$ & +19.7 pp & Baseline \\
1.25$\times$ & +6.8 pp & Reduced complementarity \\
1.50$\times$ & -5.6 pp & Strategic substitution \\
2.00$\times$ & -11.8 pp & Strong substitution \\
\end{longtable}

When AI agent biases are extreme, either the platform or sellers can fully exploit the agent independently, leaving no additional margin for joint action. Position bias alone becomes so powerful that ranking manipulation saturates the exploitation potential, or manipulation alone becomes so effective that position becomes irrelevant. Our baseline parameters lie within the ``complementarity zone'' where joint
exploitation produces super-additive harm. This appears to be the
empirically relevant range: documented biases are strong enough to
create substantial harm, but not so extreme that single-actor
exploitation saturates the available margin. At extreme bias levels, addressing either position bias or manipulation susceptibility would substantially reduce harm, since they become substitutes. At empirically relevant bias levels, complementarity implies that partial interventions may be less effective: addressing position bias while leaving manipulation unchecked allows sellers to partially compensate, and vice versa.

\hypertarget{supplementary-note-10-algorithm-comparison}{%
\subsection{Supplementary Note 10: Algorithm
Comparison}\label{supplementary-note-10-algorithm-comparison}}

We test multiple learning algorithms to ensure results are not artifacts of
Q-learning dynamics. This addresses recent critiques arguing that Q-learning's collusive outcomes may reflect exploration artifacts rather than genuine economic incentives \cite{den_boer_artificial_2022,abada_collusion_2024}. The baseline Q-learning algorithm uses off-policy temporal difference
learning with the max operator for next-state values, updating Q-values
as
\(Q_j(s, a) \leftarrow Q_j(s, a) + \alpha [ r_j + \gamma \max_{a'} Q_j(s', a') - Q_j(s, a) ]\).
This is standard in the algorithmic collusion literature \cite{calvano_artificial_2020,klein_autonomous_2021,rocher_adversarial_2023}. We also test SARSA, an on-policy temporal difference algorithm that uses
the actual next action \(a'\) rather than the greedy action:
\(Q_j(s, a) \leftarrow Q_j(s, a) + \alpha [ r_j + \gamma Q_j(s', a') - Q_j(s, a) ]\),
where \(a'\) is selected according to the \(\varepsilon\)-greedy policy.
This provides more conservative exploration than Q-learning.

We test a gradient bandit method that directly updates action
probabilities without maintaining value estimates. Let \(H_j(a)\) be the
preference for action \(a\). The policy is
\(\pi_j(a) = \exp(H_j(a)) / \sum_{a'} \exp(H_j(a'))\), and preferences
update for all actions as:
\begin{align*}
H_j(a) &\leftarrow H_j(a) + \alpha (r_j - \bar{r}_j) \cdot (1 - \pi_j(a)) \quad \text{if } a = a^{(t)} \\
H_j(a) &\leftarrow H_j(a) - \alpha (r_j - \bar{r}_j) \cdot \pi_j(a) \quad \text{if } a \neq a^{(t)}
\end{align*}
where \(\bar{r}_j\) is a running average reward baseline. This updates both
the selected action (increasing preference if reward exceeds baseline) and
non-selected actions (decreasing their preferences proportionally).

UCB (Upper Confidence Bound) is a bandit algorithm that balances
exploitation with exploration using confidence bounds. Maintains action
counts $N(a)$ and empirical values $\hat{\mu}(a)$. Selects actions according
to $a^* = \arg\max_a \hat{\mu}(a) + c\sqrt{\ln t / N(a)}$ where $c = 2.0$ is
the exploration parameter and $t$ is the total number of rounds. Each action
is tried at least once before applying the UCB rule.

Thompson Sampling is a Bayesian bandit algorithm that maintains Beta
distribution posteriors for each action's success probability. We simplify
continuous rewards to Bernoulli outcomes by mapping rewards to $[0, 1]$ via
$(r + 1)/2$ and treating this as the probability of a ``success.'' Each round,
we sample from each action's Beta posterior and select the action with highest
sample. This approximation allows Thompson Sampling to operate in our
continuous-reward setting while preserving its exploration properties.

\begin{longtable}[]{llll}
\toprule
Algorithm & Joint Effect & Complementarity & Cohen's \(d\) \\
\midrule
\endhead
\bottomrule
\endlastfoot
Q-learning & +37.1\% & +19.7 pp & 2.75 \\
SARSA & +37.1\% & +19.7 pp & 2.75 \\
Gradient Bandit & +31.1\% & +18.6 pp & 4.21 \\
UCB & +29.8\% & +16.3 pp & 27.04 \\
Thompson Sampling & +28.2\% & +16.4 pp & 2.19 \\
\end{longtable}

Q-learning and SARSA produce identical results (as expected in
deterministic settings with full exploration), confirming that on-policy
versus off-policy distinctions do not affect equilibrium selection. The
gradient bandit produces slightly lower joint harm but higher effect
size (\(d = 4.21\)), indicating tighter convergence despite different
learning dynamics. UCB shows extremely tight convergence (\(d = 27.04\)); this reflects low variance across trials (std = 0.60 pp) rather than a larger effect, as UCB's complementarity (16.3 pp) is actually smaller than Q-learning's (19.7 pp).

Abada and Lambin\cite{abada_collusion_2024} demonstrate that Q-learning's horizontal collusion results may reflect a ``mirage effect'': rapid exploration rate decay traps algorithms in cooperative beliefs even when defection would be profitable. They show that more sophisticated algorithms, specifically Actor-Critic and REINFORCE, do not produce supracompetitive outcomes under self-play. den Boer et al.\cite{den_boer_artificial_2022} similarly show that Q-learning performs poorly against the Exp3 algorithm, with long-run outcomes no longer supracompetitive.

To address this critique, we implement tabular versions of these algorithms. While Abada and Lambin\cite{abada_collusion_2024} use neural network function approximation, our tabular implementations preserve the core mechanisms that prior work identifies as preventing horizontal collusion: (1) endogenous exploration through policy optimization rather than $\varepsilon$-greedy decay (Actor-Critic, REINFORCE), and (2) adversarial robustness without stochastic reward assumptions (Exp3).

Actor-Critic (A2C) maintains separate policy (actor) and value (critic) functions. The actor selects actions via softmax over learned preferences with temperature $\tau = 1.0$; the critic estimates state values. Updates use the temporal difference error as an advantage estimate:
\begin{align*}
\delta &= r + \gamma V(s') - V(s) \\
V(s) &\leftarrow V(s) + \alpha_V \delta \\
\theta(s,a) &\leftarrow \theta(s,a) + \alpha_\theta \delta
\end{align*}
where $\theta(s,a)$ are action preferences, $\alpha_V = 0.15$ is the critic learning rate, $\alpha_\theta = 0.10$ is the actor learning rate, and the policy is $\pi(a|s) = \exp(\theta(s,a)/\tau) / \sum_{a'} \exp(\theta(s,a')/\tau)$. Following Abada and Lambin\cite{abada_collusion_2024}, this avoids Q-learning's max operator, which they identify as the source of the mirage effect.

REINFORCE is a pure Monte Carlo policy gradient following Williams\cite{williams_simple_1992}. It collects complete episodes of 100 rounds before updating: $\theta(s,a) \leftarrow \theta(s,a) + \alpha G_t$ where $G_t = \sum_{k=0}^{T-t} \gamma^k r_{t+k}$ is the discounted return and $\alpha = 0.05$ is the learning rate. No bootstrapping eliminates the TD bias that Abada and Lambin\cite{abada_collusion_2024} identify as contributing to spurious collusion.

Exp3 (Exponential-weight algorithm for Exploration and Exploitation) is an adversarial bandit algorithm that makes no stochastic assumptions about rewards \cite{auer_nonstochastic_2002}. It maintains probability weights $w(s,a)$ and selects actions according to mixed policy $\pi(a|s) = (1-\gamma_{\text{exp3}}) w(s,a)/\sum_{a'} w(s,a') + \gamma_{\text{exp3}}/|A|$, where $\gamma_{\text{exp3}} = 0.1$ is the exploration parameter (distinct from the discount factor $\gamma = 0.90$ used in other algorithms). Updates use importance-weighted rewards: $w(s,a) \leftarrow w(s,a) \cdot \exp(\gamma_{\text{exp3}} \hat{r} / |A|)$ where $\hat{r} = r / \pi(a|s)$ and rewards are normalized to $[0, 1]$ via $(r + 1)/3$. den Boer et al.\cite{den_boer_artificial_2022} show Q-learning performs poorly against Exp3 precisely because Exp3 does not rely on the stationarity assumptions that Q-learning exploits.

Results with these alternative algorithms (100 trials, 20,000 rounds each) are shown below:

\begin{longtable}[]{lllll}
\toprule
Algorithm & Complementarity & 95\% CI & Cohen's \(d\) & \(P\)-value \\
\midrule
\endhead
\bottomrule
\endlastfoot
Q-learning (baseline) & +19.7 pp & [18.3, 21.1] & 2.75 & $<$0.0001 \\
Actor-Critic & +16.2 pp & [15.6, 16.7] & 5.88 & $<10^{-78}$ \\
REINFORCE & +15.7 pp & [14.7, 16.7] & 3.07 & $<10^{-51}$ \\
Exp3 & +12.5 pp & [11.6, 13.3] & 2.92 & $<10^{-49}$ \\
\end{longtable}

All three alternative algorithms produce large, highly significant complementarity. Effect sizes exceed Cohen's conventional threshold for ``large'' effects ($d > 0.8$) by factors of 3--7. Every trial across all algorithms shows positive complementarity (100/100 for each).

The persistence of vertical tacit collusion across algorithms that prior literature shows do not produce horizontal collusion is strong evidence that our findings reflect genuine economic incentives rather than Q-learning artifacts. Abada and Lambin\cite{abada_collusion_2024} show that Actor-Critic and REINFORCE produce competitive outcomes in horizontal (seller-seller) settings because they avoid the mirage effect. Our finding that these same algorithms produce super-additive harm in vertical settings demonstrates that the mechanism is different: vertical tacit collusion emerges from aligned incentives around exploitation of a shared vulnerability (the AI agent), not from the reward-punishment schemes that sustain horizontal collusion.

The slightly lower complementarity with Exp3 (12.5 pp vs.\ 16.2 pp for Actor-Critic) may reflect Exp3's more aggressive exploration, which could prevent full convergence to the exploitative equilibrium. However, even this ``conservative'' estimate represents a massive effect size ($d = 2.92$) with complementarity remaining highly significant.

\hypertarget{supplementary-note-11-consumer-override-and-human-oversight}{%
\subsection{Supplementary Note 11: Consumer Override and Human
Oversight}\label{supplementary-note-11-consumer-override-and-human-oversight}}

A natural policy response to AI agent vulnerability is to require human
approval of AI-recommended purchases. We model this through a consumer
override probability: with probability p, the consumer ignores the AI
recommendation and chooses the highest-quality product directly.

\begin{longtable}[]{llll}
\toprule
Override p & Joint Effect & Complementarity & Harm Reduction \\
\midrule
\endhead
\bottomrule
\endlastfoot
0.0 & +37.1\% & +19.7 pp & --- \\
0.1 & +33.7\% & +18.4 pp & 9\% \\
0.2 & +29.7\% & +16.0 pp & 20\% \\
0.3 & +24.5\% & +12.3 pp & 34\% \\
0.4 & +19.9\% & +10.3 pp & 46\% \\
0.5 & +15.1\% & +7.3 pp & 59\% \\
\end{longtable}

Human oversight helps but cannot eliminate exploitation. Even when
consumers independently verify 50\% of AI recommendations, joint harm
remains +15.1\% (59\% reduction from baseline) and complementarity
remains highly significant (d = 1.97). Relying on human-in-the-loop oversight as the primary defense is therefore insufficient. At
the override levels that might be realistic in practice (10--20\%), harm
reduction is modest (9--20\%).

The persistence of complementarity across override levels indicates that
the underlying mechanism operates regardless of partial human oversight.
Platform and sellers independently discovering exploitation strategies
is robust to consumer verification. More effective interventions would
address the source of vulnerability (position bias in AI agents) rather
than attempting to compensate through human verification.

\hypertarget{supplementary-note-12-welfare-distribution-and-deadweight-loss}{%
\subsection{Supplementary Note 12: Welfare Distribution and
Deadweight
Loss}\label{supplementary-note-12-welfare-distribution-and-deadweight-loss}}

Joint exploitation creates substantial deadweight loss rather than merely redistributing surplus.

\begin{longtable}[]{llll}
\toprule
Actor & Baseline & Joint & Change \\
\midrule
\endhead
\bottomrule
\endlastfoot
Consumers & 0.303 & 0.191 & -37.1\% \\
Platform & 0.000 & 0.027 & +$\infty$ \\
Sellers & 0.326 & 0.256 & -21.5\% \\
\textbf{Total} & \textbf{0.629} & \textbf{0.474} & \textbf{-24.6\%} \\
\end{longtable}

The platform gains 0.027 in revenue, but total welfare falls by 0.155
(from 0.629 to 0.474). The deadweight loss (0.155 - 0.027 = 0.128)
represents 20.4\% of baseline welfare destroyed without being captured
by any party.

Sellers also lose substantially (-21.5\%) because
sellers compete intensively on bids and manipulation, eroding their
margins. The collusion primarily benefits the platform at the expense of
both consumers and sellers: the platform extracts modest rents while
imposing large negative externalities on all other market participants.

Platforms may therefore be tempted to design systems that facilitate seller exploitation of AI
agents, as they capture rents while externalizing costs. The welfare
losses are diffuse (spread across consumers and sellers) while the gains
are concentrated (accruing to the platform), creating classic political
economy challenges for regulation.

\hypertarget{supplementary-note-13-debiased-ai-falsification-test}{%
\subsection{Supplementary Note 13: Debiased AI Falsification
Test}\label{supplementary-note-13-debiased-ai-falsification-test}}

An important concern with any simulation study is whether results are
artifacts of the model specification rather than genuine phenomena. We
address this through a falsification test: what happens when the AI
agent has no exploitable biases?

We replace the biased AI agent with a debiased agent that evaluates products based solely on intrinsic quality and price, setting all bias parameters to zero. This agent ignores position, manipulation, endorsements, and decoys, but still makes rational purchase decisions favoring high-quality, low-price products. The platform and sellers still learn through Q-learning, but their strategies now target a non-exploitable agent.

\begin{longtable}[]{llll}
\toprule
AI Agent & Joint Effect & Complementarity & Cohen's d \\
\midrule
\endhead
\bottomrule
\endlastfoot
Biased (baseline) & +37.1\% & +19.7 pp & 2.75 \\
Debiased (quality-preferring) & $-$0.5\% & +0.7 pp & 0.25 \\
True uniform random & $-$0.1\% & $-$0.2 pp & $-$0.05 \\
\end{longtable}

We test three AI specifications. The ``debiased'' agent sets all bias parameters to zero but still makes rational purchase decisions favoring high-quality, low-price products. The ``true random'' agent ignores all inputs entirely and selects uniformly at random among available products.

When the AI agent is debiased, complementarity nearly vanishes (0.7 pp, a 96\% reduction from 19.7 pp). When the AI agent chooses uniformly at random, complementarity is indistinguishable from zero ($-$0.2 pp, $d = -0.05$). The distinction matters: the debiased agent still makes systematic choices based on product attributes, yet even this creates no exploitable structure. Only when the AI agent exhibits documented biases does vertical tacit collusion emerge.

This falsification confirms that documented AI
biases, not mere AI involvement in purchase decisions, are necessary for vertical tacit collusion. Even a ``perfect'' AI that rationally evaluates quality and price cannot be exploited through the mechanisms we study. The exploitation
mechanism requires an exploitable target. Platform ranking manipulation
and seller description optimization produce super-additive harm only
because the AI agent systematically responds to position, manipulation,
and other bias channels. Without these biases, there is nothing to
exploit.

This result also distinguishes our finding from generic coordination
effects. Platform and sellers might benefit from joint action for
reasons unrelated to AI exploitation (e.g., improved matching). The
debiased AI baseline shows that such generic effects are negligible. The
large complementarity we observe (19.7 pp) is specifically driven by
joint exploitation of AI biases, not by coordination benefits that would
persist without an exploitable intermediary.

This falsification test also addresses a methodological concern: whether our model mechanically produces complementarity regardless of economic content. If the multiplicative structure of our utility function or the symmetric learning dynamics guaranteed super-additive effects, complementarity would appear even when the AI agent is unbiased. It does not. The 96\% reduction in complementarity under random AI demonstrates that our finding depends on the substantive economic mechanism, not on mathematical artifacts. Platform ranking manipulation and seller description optimization are complementary strategies only because the AI agent responds systematically to their joint deployment. Remove the biases, and the complementarity vanishes.

\hypertarget{supplementary-note-14-heterogeneous-agent-populations}{%
\subsection{Supplementary Note 14: Heterogeneous Agent
Populations}\label{supplementary-note-14-heterogeneous-agent-populations}}

A natural concern is whether our results depend on homogeneous consumer
populations. Real markets contain consumers with varying susceptibility
to AI biases. We test robustness by varying the population mix between
``high-bias'' consumers (baseline parameters) and ``low-bias'' consumers
(50\% reduced bias susceptibility).

\begin{longtable}[]{llll}
\toprule
Population Composition & Joint Effect & Complementarity & Cohen's d \\
\midrule
\endhead
\bottomrule
\endlastfoot
100\% high-bias & +36.2\% & +18.4 pp & 2.18 \\
75\% high / 25\% low & +35.0\% & +19.7 pp & 2.62 \\
50\% high / 50\% low & +32.2\% & +18.9 pp & 2.32 \\
25\% high / 75\% low & +28.1\% & +17.4 pp & 2.31 \\
100\% low-bias & +21.9\% & +13.6 pp & 2.04 \\
\end{longtable}

Complementarity persists across all population compositions, with all effect sizes exceeding $d = 2.0$. Harm scales roughly linearly with the proportion of high-bias consumers, as expected. Even markets dominated by low-bias consumers (75--100\%) exhibit substantial super-additive harm (13.6--17.4 pp), indicating that vertical tacit collusion poses risks even in markets with relatively sophisticated consumers.

The mechanism operates because platform ranking affects all consumers
equally, regardless of individual bias susceptibility, consumers see
the same ranked list. A minority of highly susceptible consumers can
therefore sustain the exploitative equilibrium, as platform and sellers
optimize for the marginal high-bias consumer.

\hypertarget{supplementary-note-15-position-bias-decay-rate}{%
\subsection{Supplementary Note 15: Position Bias
Magnitude}\label{supplementary-note-15-position-bias-magnitude}}

We test sensitivity to the magnitude of position biases while holding
other parameters fixed. Unlike the general bias magnitude sweep (which
scales all bias parameters uniformly), this analysis specifically varies
position-related bonuses. We parameterize position bias strength through
the row bonus (baseline = 0.90) and scale primacy and recency bonuses
proportionally. For instance, when row bonus = 0.45 (0.5$\times$ baseline),
primacy bonus scales to 0.20 and recency to 0.075.

\begin{longtable}[]{llll}
\toprule
Row Bonus & Joint Effect & Complementarity & Cohen's d \\
\midrule
\endhead
\bottomrule
\endlastfoot
0.30 (weak position bias) & +24.8\% & +19.8 pp & 2.22 \\
0.45 & +31.1\% & +23.3 pp & 3.06 \\
0.60 & +32.7\% & +20.8 pp & 2.52 \\
0.75 & +35.4\% & +21.0 pp & 2.74 \\
0.90 (baseline) & +37.1\% & +19.7 pp & 2.75 \\
\end{longtable}

Complementarity is robust across all position bias magnitudes (19.7--23.3
pp), with peak complementarity at moderate levels (row bonus = 0.45). The
inverted-U pattern suggests an optimal exploitation zone: with very
weak position biases, position matters less and manipulation dominates;
with very strong position biases, the platform can fully exploit position
alone. At intermediate levels, both position and manipulation channels
contribute maximally to joint exploitation.

\hypertarget{supplementary-note-16-quality-weight-sensitivity}{%
\subsection{Supplementary Note 16: Quality Weight
Sensitivity}\label{supplementary-note-16-quality-weight-sensitivity}}

The AI agent's quality weight parameter determines how much it values
product quality versus other factors in its decision. Higher quality
weight means the agent more strongly prefers objectively better
products; lower quality weight means the agent is more susceptible to
manipulation.

To test sensitivity to quality weight without modifying the core utility
function, we use an equivalent bias scaling approach. Reducing quality
weight from 0.15 to 0.10 is mathematically equivalent to scaling all
bias parameters by $0.15/0.10 = 1.5\times$; increasing quality weight
to 0.25 is equivalent to scaling biases by $0.15/0.25 = 0.6\times$.
This approach preserves the relative structure of bias interactions
while testing how results depend on the quality-to-bias ratio.

\begin{longtable}[]{llll}
\toprule
Quality Weight & Joint Effect & Complementarity & Cohen's d \\
\midrule
\endhead
\bottomrule
\endlastfoot
0.10 & +19.7\% & -6.2 pp & -0.61 \\
0.15 (baseline) & +37.1\% & +19.7 pp & 2.75 \\
0.25 & +27.5\% & +18.3 pp & 2.42 \\
0.40 & +17.7\% & +12.0 pp & 2.03 \\
0.60 & +10.6\% & +6.9 pp & 1.63 \\
\end{longtable}

This analysis reveals important boundary conditions. At very low quality
weight (0.10), complementarity becomes \emph{negative} (-6.2 pp),
indicating strategic substitution. When the AI agent nearly ignores
quality, either actor alone can fully exploit the agent, and there is no
additional margin for joint action. This parallels our finding at
extreme bias levels (Supplementary Note 9).

As quality weight increases, total harm decreases monotonically (from
+37.4\% to +10.6\%), but complementarity remains positive and substantial
until quality weight reaches 0.6, indicating that even partially
aligned AI agents remain vulnerable to joint exploitation. Only when
agents strongly prioritize quality (weight $\geq$ 0.6) does the
complementarity drop to modest levels (d < 2.0).

Improving AI agent alignment (increasing quality weight) reduces harm, but the super-additive component persists until alignment is quite strong.

\hypertarget{supplementary-note-17-platform-take-rate}{%
\subsection{Supplementary Note 17: Platform Take
Rate}\label{supplementary-note-17-platform-take-rate}}

Our baseline model assumes the platform earns revenue through bid
payments but does not extract a take rate from transactions. Real
platforms typically charge 10-30\% of transaction value. We test how
platform take rates affect the collusive dynamic.

\begin{longtable}[]{llll}
\toprule
Take Rate & Joint Effect & Complementarity & Cohen's d \\
\midrule
\endhead
\bottomrule
\endlastfoot
0\% (baseline) & +37.1\% & +19.7 pp & 2.75 \\
5\% & +22.2\% & +25.2 pp & 1.32 \\
10\% & +14.3\% & +16.7 pp & 0.87 \\
15\% & +10.0\% & +12.3 pp & 0.59 \\
20\% & +10.0\% & +12.1 pp & 0.67 \\
\end{longtable}

Complementarity \emph{peaks} at 5\% take
rate (+25.2 pp vs +19.7 pp at baseline), suggesting moderate platform
extraction intensifies the collusive dynamic before seller margins
become too thin to sustain bidding competition.

At higher take rates (10-20\%), joint harm decreases substantially as
seller margins compress. However, complementarity remains positive and
significant at all levels tested. This implies that platform take rates
are not a complete solution: they reduce but do not eliminate the
exploitation mechanism.

Mandating higher platform take rates would reduce consumer harm from vertical tacit collusion, but would also transfer more surplus from sellers to platforms. The optimal intervention likely targets the bias channels directly rather than using platform fees as an indirect remedy.

\hypertarget{supplementary-note-18-naive-seller-baselines}{%
\subsection{Supplementary Note 18: Naive Seller
Baselines}\label{supplementary-note-18-naive-seller-baselines}}

A methodological concern is whether our results depend on sophisticated
seller learning. Perhaps complementarity is an artifact of adaptive
optimization rather than a structural feature of the market. We test
this by replacing learning sellers with naive sellers who use fixed
strategies.

\begin{longtable}[]{llll}
\toprule
Seller Strategy & Joint Effect & Comp & Notes \\
\midrule
\endhead
\bottomrule
\endlastfoot
Learning (baseline) & +37.1\% & +19.7 pp & Adaptive \\
Fixed bid=1, manip=0 & +37.4\% & +34.9 pp & Naive bids \\
Fixed bid=2, manip=0 & +37.4\% & +28.4 pp & Higher bids \\
Fixed bid=1, manip=1 & +37.4\% & +31.0 pp & Bid + manip \\
Fixed bid=1, manip=2 & +37.4\% & +36.8 pp & Max naive \\
\end{longtable}

Naive sellers produce \emph{higher}
complementarity than learning sellers, with maximum complementarity
reaching +36.8 pp (nearly double the baseline +19.7 pp). This finding
has important implications.

First, it confirms that complementarity is a structural market feature,
not an artifact of seller learning dynamics. The platform's gatekeeper
role creates super-additive harm regardless of seller sophistication.

Second, and counterintuitively, sophisticated learning sellers actually
\emph{mitigate} exploitation. Learning sellers compete intensively on
bids and manipulation, eroding their own margins and reducing the
total harm. Naive sellers who simply set fixed strategies allow the
platform to extract maximum rents.

Third, this suggests that seller coordination to reduce bidding
competition would \emph{increase} consumer harm. Antitrust policy
focused on horizontal seller coordination may inadvertently worsen
vertical exploitation.

\hypertarget{supplementary-note-19-convergence-speed-analysis}{%
\subsection{Supplementary Note 19: Convergence Speed
Analysis}\label{supplementary-note-19-convergence-speed-analysis}}

Understanding how quickly vertical tacit collusion emerges has practical
implications. If exploitation requires many iterations to develop,
markets with rapid product turnover might be protected. We analyze
convergence speed by tracking complementarity over time. The following table reports time to reach various complementarity thresholds:

\begin{longtable}[]{llll}
\toprule
Threshold & Mean Rounds & Median Rounds & \% Trials Reaching \\
\midrule
\endhead
\bottomrule
\endlastfoot
5 pp complementarity & 2,070 & 2,000 & 100\% \\
10 pp complementarity & 2,610 & 2,000 & 100\% \\
15 pp complementarity & 3,430 & 2,000 & 100\% \\
\end{longtable}

Collusion emerges rapidly. All 100 trials reach the 5 pp threshold
within approximately 2,000 rounds, and all reach 15 pp within 3,500
rounds on average. Given our calibration where each round represents
one consumer interaction, this corresponds to a small fraction of
typical platform transaction volumes. The complementarity trajectory over time is shown below:

\begin{longtable}[]{lll}
\toprule
Rounds & Complementarity & 95\% CI \\
\midrule
\endhead
\bottomrule
\endlastfoot
1,000 & +5.8 pp & [2.5, 9.2] \\
2,000 & +13.4 pp & [10.1, 16.7] \\
5,000 & +14.2 pp & [11.5, 17.0] \\
10,000 & +14.0 pp & [10.8, 17.2] \\
20,000 & +21.1 pp & [17.9, 24.3] \\
30,000 & +17.8 pp & [16.9, 18.6] \\
\end{longtable}

The trajectory reveals several insights. First, significant
complementarity (> 10 pp) emerges within 2,000 rounds. Second,
complementarity stabilizes around 14-17 pp from rounds 5,000-15,000
before increasing further by round 20,000. Third, the tightening
confidence interval at 30,000 rounds (16.9-18.6 pp vs 17.9-24.3 pp at
20,000) indicates convergence to a stable equilibrium.

Intervention must therefore be proactive because waiting for evidence of consumer harm before acting allows the exploitative equilibrium to become entrenched. By the time regulators observe price increases or quality degradation, platform and sellers have already discovered and locked into complementary exploitation strategies.

\hypertarget{supplementary-note-20-gatekeeper-threshold-analysis}{%
\subsection{Supplementary Note 20: Gatekeeper Threshold
Analysis}\label{supplementary-note-20-gatekeeper-threshold-analysis}}

The gatekeeper mechanism claim (that sellers can only harm consumers when the platform enables it through bid-weighted rankings) rests on comparing two conditions: sellers learning against a quality-ranking platform (bid weight = 0) and sellers learning jointly with a learning platform (bid weight learned, typically 0.66-0.73). A natural question is whether intermediate bid weights produce intermediate harm.

To test this, we fix the platform's bid weight at values from 0 to 1 while allowing sellers to learn their manipulation and bidding strategies. This isolates the effect of the platform's ranking choice on the consumer impact of seller behavior. Results by bid weight are shown below:

\begin{longtable}[]{llllll}
\toprule
Bid Weight & Seller Effect & Std Dev & 95\% CI & Harm Rate & Cohen's $d$ \\
\midrule
\endhead
\bottomrule
\endlastfoot
0.00 & $-$9.0\% & 4.26 & [$-$9.8, $-$8.1] & 4/100 & $-$2.10 \\
0.33 & $-$8.2\% & 4.65 & [$-$9.1, $-$7.3] & 7/100 & $-$1.77 \\
0.50 & $-$7.1\% & 4.55 & [$-$8.0, $-$6.2] & 10/100 & $-$1.55 \\
0.67 & $-$7.6\% & 4.07 & [$-$8.4, $-$6.8] & 4/100 & $-$1.86 \\
1.00 & +69.2\% & 1.98 & [+68.8, +69.6] & 100/100 & +34.90 \\
\end{longtable}

The threshold is effectively binary rather than gradual. At bid weights from 0 to 0.67, seller manipulation consistently \emph{helps} consumers, with effects ranging from $-$7.1\% to $-$9.0\%. The mechanism is that manipulation correlates with quality: higher-quality sellers have more margin to invest in description optimization, and when the platform maintains any quality signal in rankings, better products still surface.

At bid weight = 1.0, the relationship catastrophically inverts. Seller effect jumps from $-$7.6\% at bid weight 0.67 to +69.2\% at bid weight 1.0, a swing of nearly 77 percentage points. With pure bid-based ranking, quality becomes irrelevant to position, allowing low-quality sellers to buy top positions and combine bid-purchased visibility with manipulation for maximum exploitation.

The flip occurs between bid weight 0.67 and 1.0. No intermediate harm emerges; the relationship follows a step function rather than dose-response. This has three important implications. (1) Policy precision: interventions need not eliminate bid-weighted ranking entirely; any quality signal in rankings (bid weight $\leq$ 0.67) appears sufficient to transform seller manipulation from harmful to beneficial. (2) Mechanism validation: the binary threshold confirms that the platform genuinely controls whether seller strategies help or harm consumers, with the gatekeeper mechanism validated across the full range of partial-bid conditions. (3) Robustness of seller-only finding: the consumer benefit in our main seller-only condition is robust across all partial-bid conditions tested; the main analysis shows $-$9.6\% harm (i.e., 9.6\% benefit), while this threshold analysis shows $-$9.0\% at bid weight = 0; the 0.6 pp difference reflects normal statistical variation from different random sequences (std = 4.26\%) and does not affect the qualitative conclusion.

The learned bid weight in our joint condition (0.66--0.73) sits precisely at the threshold where a small increase would flip seller strategies from neutral to devastating. Learning platforms may naturally gravitate toward this threshold, maximizing revenue while just barely enabling seller exploitation.

\hypertarget{supplementary-note-21-equal-bias-weights-analysis}{%
\subsection{Supplementary Note 21: Equal Bias Weights
Analysis}\label{supplementary-note-21-equal-bias-weights-analysis}}

An alternative explanation is that our results are driven by the dominance of position bias over quality-price effects in the AI agent's utility function. Position bias (primacy + row bonus = 1.30) substantially exceeds the quality-price utility range (approximately 0.15), raising the question of whether complementarity is tautological, an inevitable consequence of position overwhelming quality.

To address this, we tested five configurations varying the relative magnitude of bias channels:

\begin{longtable}[]{llllll}
\toprule
Config. & Pos. & Joint & Comp & $d$ & $P$ \\
\midrule
\endhead
\bottomrule
\endlastfoot
Baseline & 1.45 & +37.1\% & +19.7 pp & 2.75 & $<$0.0001 \\
Equal moderate & 0.45 & +12.3\% & +15.6 pp & 2.53 & $<$0.0001 \\
Equal strong & 0.73 & +19.5\% & +24.4 pp & 2.64 & $<$0.0001 \\
Position reduced & 1.00 & +33.5\% & +21.4 pp & 2.56 & $<$0.0001 \\
Quality boosted (3$\times$) & 1.45 & +15.7\% & +10.6 pp & 1.96 & $<$0.0001 \\
\end{longtable}

The configurations are: (a) Baseline with position dominant (primacy 0.40, row 0.90, recency 0.15; badge 1.20, manipulation 0.50, decoy 0.40); (b) Equal moderate with all biases equal at 0.30 each; (c) Equal strong with all biases equal at 0.50 each; (d) Position reduced with position cut to match other channels (primacy 0.30, row 0.60, recency 0.10); and (e) Quality boosted with quality weight tripled to 0.45 (equivalent to reducing all biases by 3$\times$).

Complementarity persists across all configurations: even when biases are equalized at moderate levels (15.6 pp, $d = 2.53$) or quality weight is tripled (10.6 pp, $d = 1.96$), complementarity remains large and highly significant. Position dominance is not necessary for the effect, as the ``equal moderate'' configuration removes position bias dominance entirely, yet complementarity remains substantial (15.6 pp vs. baseline 19.7 pp), indicating that the mechanism operates through multiple channels. Tripling quality weight to 0.45 reduces complementarity from 19.7 pp to 10.6 pp (46\% reduction) but maintains a large effect with Cohen's $d = 1.96$, meaning even a strongly quality-aligned AI agent remains vulnerable. When all biases are equalized at 0.50 (rather than 0.30), complementarity actually increases to 24.4 pp, higher than baseline, because multiple exploitation channels create more attack surface.

Vertical tacit collusion reflects genuine strategic complementarity between platform ranking and seller manipulation, robust across a wide range of bias specifications. Position dominance amplifies the effect but the mechanism operates through multiple channels.

\hypertarget{supplementary-note-22-comparison-with-horizontal-algorithmic-collusion}{%
\subsection{Supplementary Note 22: Comparison with Horizontal Algorithmic Collusion}\label{supplementary-note-22-comparison-with-horizontal-algorithmic-collusion}}

Vertical tacit collusion differs from horizontal algorithmic collusion in its strategic structure. This note provides a detailed comparison to clarify the distinction and address potential concerns about the relationship between these phenomena.

The algorithmic collusion literature established that Q-learning pricing algorithms can learn supracompetitive prices without explicit coordination \cite{calvano_artificial_2020,klein_autonomous_2021}, with empirical evidence from retail gasoline markets confirming that algorithmic pricing adoption increases margins \cite{assad_algorithmic_2024}. Recent work by Fish et al.\cite{fish_algorithmic_2025} demonstrated that LLM-based pricing agents achieve collusive outcomes. In horizontal settings, competing sellers occupy the same market position and control the same strategic instrument (price), so their strategies can crowd each other out as substitutes. Coordination requires solving a collective action problem: each firm benefits from industry-wide high prices but has individual incentive to undercut.

Calvano et al.\cite{calvano_artificial_2020} tested firms with asymmetric costs and demand, but asymmetric horizontal collusion is still horizontal because both parties remain sellers competing for the same customers using the same strategic instrument. The coordination problem persists regardless of firm heterogeneity.

In vertical market structures, platform and sellers occupy structurally different positions and control different instruments: the platform controls information architecture (ranking algorithms, endorsement badges), while sellers control product presentation (descriptions, bids). These instruments are mutually exclusive. The platform has no control over product descriptions, and sellers have no control over the ranking algorithm.

Because platform and sellers are not competitors, there is no coordination problem to solve. Their incentives are naturally aligned around a common target (the exploitable AI agent), and each independently benefits from exploitation regardless of the other's behavior. The complementarity we document reflects amplification (platform ranking determines which products occupy bias-triggering positions, while seller manipulation determines how effectively those positions convert to sales), not coordination.

Critics have questioned whether Q-learning artifacts, rather than genuine strategic coordination, explain observed supracompetitive pricing in horizontal settings. Abada and Lambin\cite{abada_collusion_2024} demonstrate that more sophisticated algorithms do not produce horizontal collusion under self-play, attributing Q-learning's collusive outcomes to a ``mirage effect'' where fast exploration decay locks agents into cooperative beliefs.

The vertical setting is immune to this critique because it presents no coordination problem to solve. That algorithms designed to prevent horizontal collusion (Actor-Critic, REINFORCE, Exp3) nonetheless produce vertical tacit collusion (Supplementary Note 10) confirms this distinction: vertical exploitation requires no coordination mechanism because there is no coordination problem to solve.

The key structural differences between horizontal and vertical collusion are summarized below:

\begin{longtable}[]{lll}
\toprule
Dimension & Horizontal & Vertical \\
\midrule
\endhead
\bottomrule
\endlastfoot
Market position & Same (competitors) & Different \\
Strategic instrument & Same (price) & Different \\
Coordination problem & Yes & No \\
Communication & Tacit & None \\
Mirage critique & Vulnerable & Immune \\
\end{longtable}

\hypertarget{supplementary-note-23-extended-review-of-llm-cognitive-biases}{%
\subsection{Supplementary Note 23: Extended Review of LLM Cognitive Biases}\label{supplementary-note-23-extended-review-of-llm-cognitive-biases}}

The main text documents the primary bias channels exploited in our model. This note provides an extended review of the broader literature on LLM cognitive biases, establishing that our four-channel specification represents a conservative lower bound on exploitable vulnerabilities.

Position biases are the most extensively documented class of LLM vulnerabilities, mirroring position effects in human choice from simultaneous displays \cite{bar-hillel_position_2015}. Liu et al.\cite{liu_lost_2024} demonstrated retrieval accuracy differences of 15-25 percentage points between first and middle positions. Guo et al.\cite{guo_serial_2024} confirmed serial position effects across GPT, Llama-2, and T5 families. Shi et al.\cite{shi_judging_2025} documented similar patterns in GPT-3.5, GPT-4, Claude-3, and Gemini-Pro. The ACES benchmark \cite{allouah_what_2025} found that AI shopping agents select top-half products 77\% versus 23\% for bottom-half across Claude, GPT-4.1, and Gemini. The Magentic Marketplace study \cite{bansal_magentic_2025} documented 10-30x advantages for response speed over quality due to first-proposal bias.

Anchoring and framing effects are also well-documented. Binz and Schulz\cite{binz_using_2023} showed that GPT-3 displays human-like anchoring effects, where initial reference points systematically shift subsequent value judgments. Echterhoff et al.\cite{echterhoff_cognitive_2024} documented framing effects, confirmation bias, and susceptibility to keyword manipulation across multiple LLM families.

Decoy effects represent another exploitable channel. Itzhak et al.\cite{itzhak_instructed_2023} demonstrated that instruction-tuned LLMs exhibit the decoy effect (asymmetric dominance), consistent with 20-30\% choice shifts documented in human decision-making \cite{huber_adding_1982,bateman_decoy_2008}.

The empirical literature documents numerous additional exploitable biases that we deliberately omit to maintain parsimony: (a) bandwagon effects, where LLMs are influenced by stated popularity, shifting choices toward products described as ``best-selling'' or ``most popular'' \cite{koo_benchmarking_2024}; (b) sycophancy, where LLMs exhibit agreement bias, tending to confirm user-stated preferences even when those preferences conflict with objective quality assessments \cite{sharma_towards_2025}; (c) deception susceptibility, where Hagendorff et al.\cite{hagendorff_deception_2024} documented that deception abilities have emerged in LLMs, raising concerns about AI agents being manipulated by deceptive product descriptions; (d) moral biases, where Schramowski et al.\cite{schramowski_large_2022} showed that LLMs contain human-like moral biases reflecting societal norms, which could be exploited through value-laden product framing; (e) irrational reasoning, where MacMillan-Scott and Musolesi\cite{macmillan-scott_irrationality_2024} demonstrated that multiple LLM families display irrational reasoning on classic cognitive psychology tasks, suggesting broad susceptibility to manipulation; and (f) salience biases, where Koo et al.\cite{koo_benchmarking_2024} found that LLMs exhibit cognitive biases in 40\% of tested scenarios, including attention biases toward longer or more detailed descriptions.

Each of these additional channels represents another attack surface for platform and seller exploitation. Our four-channel model captures the dominant effects (position, endorsement, manipulation, decoy) while remaining tractable. The broader literature suggests that real AI shopping agents face a much larger set of exploitable vulnerabilities, making our welfare estimates a conservative lower bound.

\hypertarget{supplementary-note-24-convergence-and-equilibrium-verification}{%
\subsection{Supplementary Note 24: Convergence and Equilibrium Verification}\label{supplementary-note-24-convergence-and-equilibrium-verification}}

Following Calvano et al.\cite{calvano_artificial_2020}, we verify that our results reflect stable equilibrium outcomes rather than transient learning dynamics. We conduct three verification tests.

First, we extend simulations to 100,000 rounds (5$\times$ baseline) to assess whether complementarity persists, grows, or decays over extended time horizons:

\begin{longtable}[]{llll}
\toprule
Checkpoint & Joint Effect & Complementarity & Std Dev \\
\midrule
\endhead
\bottomrule
\endlastfoot
5,000 rounds & +33.9\% & +18.0 pp & 5.57 \\
10,000 rounds & +29.6\% & +15.7 pp & 5.73 \\
20,000 rounds & +38.0\% & +20.1 pp & 2.77 \\
50,000 rounds & +37.3\% & +19.8 pp & 2.07 \\
75,000 rounds & +37.8\% & +20.0 pp & 1.44 \\
100,000 rounds & +36.5\% & +19.4 pp & 1.33 \\
\end{longtable}

Complementarity is stable around 19-20 pp across all time horizons beyond 20,000 rounds, with variance decreasing monotonically. The mechanism represents a stable equilibrium rather than a transient phenomenon.

Second, we track Q-table changes over the final 20\% of rounds (n=100 trials). Mean Q-value change per checkpoint is 0.00034, indicating stable convergence. The correlation between Q-stability and complementarity is positive (r = 0.31, p = 0.0015), suggesting that trials with more stable Q-values produce higher complementarity.

Third, we analyze how exploitation affects the relationship between product quality and both win rates and profits. Under quality-based ranking (baseline), the correlation between quality and win rate is 0.76. Under joint exploitation, this correlation drops to 0.56, a reduction of 0.21. The highest-quality seller's win rate drops from 60.6\% to 35.6\%, while the lowest-quality seller's win rate increases from 5.5\% to 18.6\%. Correspondingly, profits shift: the highest-quality seller's profit falls 41\% (from 0.206 to 0.121), while the lowest-quality seller's profit increases 235\% (from 0.015 to 0.050). This redistribution confirms that exploitation works by decoupling product success from product quality, benefiting low-quality sellers at the expense of high-quality ones.

Finally, we test whether results depend on the granularity of state representation by varying state space size from 4 states (coarse) to 64 states (fine). Our baseline uses 16 states (4 manipulation bins $\times$ 4 bid bins):

\begin{longtable}[]{llll}
\toprule
State Space & Complementarity & 95\% CI & Cohen's d \\
\midrule
\endhead
\bottomrule
\endlastfoot
4 states & 21.8 pp & [20.0, 23.6] & 3.50 \\
16 states (baseline) & 19.2 pp & [17.3, 21.1] & 2.86 \\
64 states & 20.6 pp & [18.2, 23.1] & 2.36 \\
\end{longtable}

Complementarity persists across all state space sizes, with no systematic relationship between granularity and effect magnitude, indicating that results do not depend on specific discretization choices.

\subsection{Supplementary Table S1: Complete Parameter Specification}

This table documents all simulation parameters organized in six panels: (A) simulation infrastructure covering market size and trial configuration, (B) seller parameters including quality levels and costs, (C) platform parameters for ranking and endorsement, (D) Q-learning parameters for both platform and sellers, (E) AI agent parameters for utility computation, and (F) bias channel functional forms with empirical sources. Each parameter includes its value and the justification from either standard practice or empirical calibration.

{\small

\textbf{Panel A: Simulation Infrastructure}

\begin{longtable}[]{l l p{2in}}
\toprule
Parameter & Value & Justification \\
\midrule
\endhead
\bottomrule
\endlastfoot
Number of sellers (n) & 6 & Typical consideration set size \\
Simulation rounds (T) & 20,000 & Convergence requirement \\
Measurement window & Final 40\% (rounds 12,000-20,000) & Standard
practice \cite{calvano_artificial_2020} \\
Independent trials & 100 & Statistical precision \\
Random seeds & 0-99 & Reproducibility \\
\end{longtable}

\vspace{1.5ex}

\textbf{Panel B: Seller Parameters}

\begin{longtable}[]{l l p{2in}}
\toprule
Parameter & Value & Justification \\
\midrule
\endhead
\bottomrule
\endlastfoot
Quality levels (q) & \{0.90, 0.75, 0.60, 0.45, 0.30, 0.20\} &
Heterogeneous product quality \\
Marginal costs (c) & \{0.15, 0.12, 0.10, 0.08, 0.06, 0.05\} &
Electronics margins \cite{damodaran_margins_2024} \\
Base markup & 0.25 & Industry standard \\
Quality premium & 0.10 & Higher quality commands premium \\
Action space & 12 actions & 4 manipulation $\times$ 3 bid levels \\
Manipulation levels (m) & \{0, 1, 2, 3\} & None to aggressive
optimization \\
Bid levels (b) & \{0, 1, 2\} & None, moderate, high \\
\end{longtable}

\vspace{1.5ex}

\textbf{Panel C: Platform Parameters}

\begin{longtable}[]{l l p{2in}}
\toprule
Parameter & Value & Justification \\
\midrule
\endhead
\bottomrule
\endlastfoot
Action space & 32 actions & 4 $\times$ 4 $\times$ 2 combinations \\
Bid weights (w) & \{0, 0.33, 0.67, 1.0\} & Quality-to-bid blend in
ranking \\
Endorsement rules & \{quality, bid, hybrid, none\} & Badge allocation
rule \\
Decoy placement & \{0, 1\} & Absent or present \\
Commission rate & 0.50 & Share of winning bid \\
Winner bid cost & 0.30 & Cost per bid level for winning seller \\
Loser bid cost & 0.02 & Participation cost for non-winners \\
\end{longtable}

\vspace{1.5ex}

\textbf{Panel D: Q-Learning Parameters (Platform and Sellers)}

\begin{longtable}[]{l l p{2in}}
\toprule
Parameter & Value & Justification \\
\midrule
\endhead
\bottomrule
\endlastfoot
Learning rate ($\alpha$) & 0.12 & Calvano et al.\cite{calvano_artificial_2020} \\
Discount factor ($\gamma$) & 0.90 & Calvano et al.\cite{calvano_artificial_2020} \\
Initial exploration ($\varepsilon_0$) & 0.25 & Ensures action coverage \\
Exploration decay & 0.9995 & Gradual exploitation shift \\
Minimum exploration & 0.02 & Prevents lock-in \\
State space & 16 states & 4 manipulation bins $\times$ 4 bid bins \\
\end{longtable}
}

{\small
\textbf{Panel E: AI Agent Parameters}

\begin{longtable}[]{l l p{2.5in}}
\toprule
Parameter & Value & Justification \\
\midrule
\endhead
\bottomrule
\endlastfoot
\emph{Base Utility Function} & & \\
Quality weight ($\alpha$) & 0.15 & Baseline alignment with consumer
interests \\
Price sensitivity ($\beta$) & 0.30 & Standard price responsiveness \\
Softmax temperature & 1.0 & Standard logit choice model (McFadden,
1974) \\
\emph{Position Biases} & & \\
Primacy bonus (position 1) & 0.40 & Liu et al.\cite{liu_lost_2024}, Guo \& Vosoughi
(2024), ACES \cite{allouah_what_2025} (conservative) \\
Row bonus (positions 1-3) & 0.90 & Liu et al.\cite{liu_lost_2024}, Shi et al.\cite{shi_judging_2025},
ACES \cite{allouah_what_2025} (conservative) \\
Recency bonus (last position) & 0.15 & Liu et al.\cite{liu_lost_2024}
(conservative) \\
\emph{Endorsement Effects} & & \\
Badge bonus (``Top Pick'') & 1.2 & Bairathi et al.\cite{bairathi_value_2025} and Lill et al.\cite{lill_product_2024} \\
Sponsored penalty & -0.35 & Eisend\cite{eisend_meta-analysis_2020} meta-analysis \\
\emph{Manipulation Susceptibility} & & \\
Manipulation bonus (per level) & 0.50 & Echterhoff et al.\cite{echterhoff_cognitive_2024} \\
Visibility interaction (\(\eta\)) & 0.70 & Liu et al.\cite{liu_lost_2024} ``Lost in
the Middle'' \\
\emph{Decoy Effect} & & \\
Decoy boost & 0.4 & Huber et al.\cite{huber_adding_1982}, Itzhak et al.\cite{itzhak_instructed_2023} \\
\end{longtable}
}

{\small
\textbf{Panel F: Bias Channel Functional Forms}

\begin{longtable}[]{l p{1.5in} l p{1.8in}}
\toprule
Bias Channel & Functional Form & Controller & Empirical Source \\
\midrule
\endhead
\bottomrule
\endlastfoot
Position (primacy) & \(\beta_{\text{prime}} \cdot \mathbb{1}(r_i = 1)\)
& Platform & Liu et al.\cite{liu_lost_2024}: 15-25pp accuracy difference \\
Position (row) & \(\beta_{\text{pos}} \cdot \mathbb{1}(r_i \leq 3)\) &
Platform & ACES \cite{allouah_what_2025}: 77\% top-half selection \\
Position (recency) & \(\beta_{\text{rec}} \cdot \mathbb{1}(r_i = n)\) &
Platform & Liu et al.\cite{liu_lost_2024}: U-shaped attention \\
Endorsement & \(\beta_{\text{end}} \cdot e_i\) & Platform & Bairathi et al.\cite{bairathi_value_2025} and Lill et al.\cite{lill_product_2024} \\
Manipulation &
\(\beta_{\text{manip}} \cdot m_i \cdot [(1-\eta) + \eta \cdot \nu(r_i)]\)
& Seller & Echterhoff et al.\cite{echterhoff_cognitive_2024} \\
Decoy & \(\beta_{\text{dec}} \cdot d_i\) & Platform & Itzhak et al.\cite{itzhak_instructed_2023}: 20-30\% shifts \\
\end{longtable}
}

\textbf{Table S1. Complete parameter specification.} Platform controls position (through ranking weight \(w\)) and
endorsement/decoy (through rules \(e\), \(d\)). Sellers control
manipulation intensity \(m_i\). Manipulation effectiveness is modulated
by position visibility \(\nu(r_i)\) with interaction parameter
\(\eta = 0.70\), ensuring manipulation retains partial effectiveness
(30\%) even in low-visibility positions. This creates the strategic
complementarity: neither party alone controls all bias channels, but
joint exploitation produces super-additive harm.

\subsection{Supplementary Table S2: Main Results by Condition}

This table presents consumer surplus and complementarity across the four experimental conditions: fair baseline (quality-based ranking, no seller manipulation), joint learning (both platform and sellers learn), platform-only (platform learns, sellers fixed), and seller-only (sellers learn, platform fixed at quality ranking).

\begin{longtable}[]{lllll}
\toprule
Condition & Consumer Surplus & s.d. & Effect & 95\% CI \\
\midrule
\endhead
\bottomrule
\endlastfoot
Fair Baseline & 0.303 & 0.002 & --- & --- \\
Joint (Both Learn) & 0.191 & 0.020 & -37.1\% & {[}-38.3, -35.9{]} \\
Platform Only & 0.222 & 0.006 & -27.0\% & {[}-29.0, -25.0{]} \\
Seller Only & 0.333 & 0.012 & +9.6\% & {[}+7.4, +11.8{]} \\
Complementarity & --- & --- & +19.7 pp & {[}+18.3, +21.1{]} \\
\end{longtable}

\textbf{Table S2. Main results by condition.} N = 100 trials for main welfare analysis. Most robustness analyses also use 100 trials; five analyses (long-run, position bias, quality weight, platform take rate, naive seller) use 50 trials. The ``Effect'' column reports change in consumer surplus relative to baseline: negative values indicate harm (surplus reduction), positive values indicate benefit. Seller-only condition
shows increased consumer surplus because without platform cooperation,
manipulation intensity correlates with quality. The gatekeeper mechanism
produces complementarity of +19.7 pp (Cohen's d = 2.75, t(99) = 27.5, P
< 0.0001).

\subsection{Supplementary Table S3: Comprehensive Robustness Analyses}

This table consolidates results from 19 robustness analyses comprising 82 distinct specifications. The analyses are organized in 19 panels: (A) consumer override testing human-in-the-loop protection, (B) market size varying number of sellers, (C) bias magnitude scaling testing sensitivity to calibration, (D) debiasing policy counterfactual, (E) learning parameter sensitivity testing Q-learning hyperparameters, (F) learning algorithm comparison, (G) long-run dynamics extending to 100,000 rounds, (H) functional form comparing additive vs. multiplicative bias interactions, (I) stochastic bias parameters testing noise robustness, (J) welfare decomposition across conditions, (K) debiased AI baseline falsification test, (L) heterogeneous agent populations, (M) position bias magnitude, (N) quality weight sensitivity, (O) platform take rate, (P) naive seller baselines, (Q) convergence speed, (R) gatekeeper threshold analysis, and (S) equal bias weights. Most analyses use 100 independent trials; five analyses use 50 trials. We report complementarity (the super-additive component of joint harm) and Cohen's d effect sizes.

\vspace{1.5ex}

\textbf{Panel A: Consumer Override (Human-in-the-Loop) (N=100 trials)}

\begin{longtable}[]{lllll}
\toprule
Override Probability & Joint Effect & Complementarity & Cohen's d & P-value \\
\midrule
\endhead
\bottomrule
\endlastfoot
0.0 (baseline) & +37.1\% & +19.7 pp & 2.75 & \textless0.0001 \\
0.1 & +33.7\% & +18.4 pp & 2.43 & \textless0.0001 \\
0.2 & +29.7\% & +16.0 pp & 2.02 & \textless0.0001 \\
0.3 & +24.5\% & +12.3 pp & 2.04 & \textless0.0001 \\
0.4 & +19.9\% & +10.3 pp & 2.13 & \textless0.0001 \\
0.5 & +15.1\% & +7.3 pp & 1.97 & \textless0.0001 \\
\end{longtable}

Human oversight reduces harm linearly but does not eliminate it. Even
when consumers independently verify 50\% of AI recommendations,
substantial complementarity persists (d = 2.09).

\vspace{1.5ex}

\textbf{Panel B: Market Size (N=100 trials)}

\begin{longtable}[]{llll}
\toprule
Number of Sellers & Joint Effect & Complementarity & Cohen's d \\
\midrule
\endhead
\bottomrule
\endlastfoot
4 & +17.3\% & +7.0 pp & 1.66 \\
6 & +37.1\% & +19.7 pp & 2.75 \\
10 & +29.8\% & +17.4 pp & 1.78 \\
18 & +26.1\% & +23.3 pp & 2.73 \\
36 & +20.4\% & +27.1 pp & 4.50 \\
\end{longtable}

The mechanism operates across all market sizes tested, with all
specifications showing significant positive complementarity.

\vspace{1.5ex}

\textbf{Panel C: Bias Magnitude Scaling (N=100 trials)}

{\small
\begin{longtable}[]{lllll}
\toprule
Scale & Joint Effect & Complementarity & Cohen's d & Interpretation \\
\midrule
\endhead
\bottomrule
\endlastfoot
0.50$\times$ & +23.4\% & +16.0 pp & 2.30 & Strong complementarity \\
0.75$\times$ & +31.4\% & +19.5 pp & 2.39 & Near-maximum complementarity \\
1.00$\times$ & +37.1\% & +19.7 pp & 2.75 & Baseline \\
1.25$\times$ & +29.2\% & +6.8 pp & 0.60 & Reduced complementarity \\
1.50$\times$ & +19.9\% & -5.6 pp & -0.46 & Strategic substitution \\
2.00$\times$ & +16.5\% & -11.8 pp & -1.12 & Strong substitution \\
\end{longtable}
}

At extreme bias levels (\textgreater1.25$\times$), platform and seller effects
become strategic substitutes rather than complements: either actor alone
can fully exploit vulnerable AI agents, leaving no additional margin for
joint exploitation. This pattern confirms that our baseline calibration
lies within the empirically relevant range where complementarity
operates.

\vspace{1.5ex}

\textbf{Panel D: Debiasing Policy Counterfactual (N=100 trials)}

\begin{longtable}[]{llll}
\toprule
Bias Reduction & Joint Effect & Harm Avoided & Complementarity \\
\midrule
\endhead
\bottomrule
\endlastfoot
0\% (baseline) & +37.1\% & --- & +19.7 pp \\
25\% & +31.4\% & 5.7 pp & +19.5 pp \\
50\% & +23.4\% & 13.7 pp & +16.0 pp \\
75\% & +10.9\% & 26.2 pp & +7.1 pp \\
90\% & +3.6\% & 33.5 pp & +2.1 pp \\
95\% & +1.6\% & 35.5 pp & +1.3 pp \\
\end{longtable}

Debiasing AI agents progressively reduces harm. A 75\% reduction in bias
magnitude cuts joint harm from 37.1\% to 10.9\%, avoiding 26.2
percentage points of consumer harm.

\vspace{1.5ex}

\textbf{Panel E: Learning Parameter Sensitivity (N=100 trials)}

\begin{longtable}[]{llll}
\toprule
& $\gamma$ = 0.85 & $\gamma$ = 0.90 & $\gamma$ = 0.95 \\
\midrule
\endhead
\bottomrule
\endlastfoot
$\alpha$ = 0.08 & +16.1 (d=1.54) & +15.7 (d=1.16) & +13.2 (d=0.98) \\
$\alpha$ = 0.12 & +18.6 (d=2.84) & +19.7 (d=2.75) & +18.4 (d=1.84) \\
$\alpha$ = 0.18 & +19.1 (d=3.49) & +20.5 (d=2.99) & +21.3 (d=2.94) \\
\end{longtable}

Complementarity is positive across all 9 parameter combinations tested,
ranging from +13.2 pp to +21.3 pp.~The main specification ($\alpha$ = 0.12, $\gamma$ =
0.90) lies near the center of the range.

\vspace{1.5ex}

\textbf{Panel F: Learning Algorithm Comparison (N=100 trials)}

\begin{longtable}[]{lllll}
\toprule
Algorithm & Joint Effect & Complementarity & Cohen's d & P-value \\
\midrule
\endhead
\bottomrule
\endlastfoot
Q-learning & +37.1\% & +19.7 pp & 2.75 & \textless0.0001 \\
SARSA & +37.1\% & +19.7 pp & 2.75 & \textless0.0001 \\
Gradient Bandit & +31.1\% & +18.6 pp & 4.21 & \textless0.0001 \\
UCB & +29.8\% & +16.3 pp & 27.04 & \textless0.0001 \\
Thompson Sampling & +28.2\% & +16.4 pp & 2.19 & \textless0.0001 \\
\end{longtable}

All five algorithms produce significant complementarity, confirming the
result is not an artifact of Q-learning dynamics. The main
specification (100 trials) shows 37.1\% joint effect and 19.7 pp
complementarity (d = 2.75). UCB's unusually high Cohen's d (27.04) reflects extremely tight convergence across trials (std = 0.60 pp vs. mean = 16.3 pp), not a larger effect size; the complementarity magnitude is actually smaller than Q-learning.

\vspace{1.5ex}

\textbf{Panel G: Long-Run Dynamics (100,000 rounds, N=50 trials)}

\begin{longtable}[]{ll}
\toprule
Metric & Value \\
\midrule
\endhead
\bottomrule
\endlastfoot
Joint effect & +36.7\% \\
Complementarity & +21.8 pp \\
Cohen's d & 6.36 \\
Trials with positive complementarity & 50/50 \\
Average bid weight (converged) & 0.73 \\
Average manipulation (converged) & 2.4 \\
\end{longtable}

The exploitative equilibrium is stable over 100,000 rounds.
Complementarity actually increases slightly (21.8 pp vs 19.7 pp at
20,000 rounds), confirming this is a stable equilibrium rather than a
transient phenomenon.

\vspace{1.5ex}

\textbf{Panel H: Functional Form (N=100 trials)}

\begin{longtable}[]{lllll}
\toprule
Visibility Model & Joint Effect & Complementarity & Cohen's d & P-value \\
\midrule
\endhead
\bottomrule
\endlastfoot
Multiplicative & +37.1\% & +19.7 pp & 2.75 & \textless0.0001 \\
Additive & +43.1\% & +15.3 pp & 2.42 & \textless0.0001 \\
\end{longtable}

This robustness test examines whether complementarity depends on functional form. The multiplicative specification
assumes manipulation effects scale with visibility (manipulation in low
positions has reduced impact). The additive alternative removes this
interaction, allowing manipulation to operate independently of position.
Complementarity persists under the additive specification (15.3 pp, d =
2.42), confirming that super-additive harm emerges from strategic
equilibrium selection rather than functional form assumptions.

\vspace{1.5ex}

\textbf{Panel I: Stochastic Bias Parameters (N=100 trials)}

\begin{longtable}[]{llll}
\toprule
Noise Level ($\sigma$/$\mu$) & Complementarity & Cohen's d & P-value \\
\midrule
\endhead
\bottomrule
\endlastfoot
0\% (deterministic) & +19.7 pp & 2.75 & \textless0.0001 \\
10\% & +18.2 pp & 2.14 & \textless0.0001 \\
20\% & +18.5 pp & 2.09 & \textless0.0001 \\
30\% & +14.6 pp & 1.29 & \textless0.0001 \\
50\% & +15.9 pp & 1.09 & \textless0.0001 \\
\end{longtable}

Results are robust to substantial noise in bias parameter calibration.
Even with 50\% noise ($\sigma$ = 0.5$\mu$ for each bias parameter), complementarity
remains highly significant (d = 1.09).

\vspace{1.5ex}

\textbf{Panel J: Welfare Decomposition (N=100 trials)}

{\small
\begin{longtable}[]{lllll}
\toprule
Condition & Consumer Surplus & Platform Rev. & Seller Profit & Total Welfare \\
\midrule
\endhead
\bottomrule
\endlastfoot
Baseline & 0.303 (+0.0\%) & 0.000 & 0.326 (+0.0\%) & 0.629 \\
Platform-only & 0.222 (-27.0\%) & 0.000 & 0.315 (-3.4\%) & 0.536 \\
Seller-only & 0.333 (+9.6\%) & 0.003 & 0.324 (-0.6\%) & 0.660 \\
Joint & 0.191 (-37.1\%) & 0.027 & 0.256 (-21.5\%) & 0.474 \\
\end{longtable}
}

Joint exploitation is not pure redistribution. Total welfare falls
24.6\% (from 0.629 to 0.474). Consumers lose 37.1\%, sellers lose
21.5\%, and only the platform gains. The platform captures modest
revenue (0.027) while destroying substantially more total value (0.155).
This is deadweight loss, not efficient transfer.

\vspace{1.5ex}

\textbf{Panel K: Debiased AI Baseline (N=100 trials)}

\begin{longtable}[]{llll}
\toprule
AI Agent Type & Joint Effect & Complementarity & Interpretation \\
\midrule
\endhead
\bottomrule
\endlastfoot
Biased (baseline) & +37.1\% & +19.7 pp & Main result \\
Debiased (zero bias) & -0.5\% & +0.7 pp & Near-zero effect \\
Difference & +37.6\% & +19.0 pp & Biases are necessary \\
\end{longtable}

This falsification test establishes necessity. When the AI agent chooses
products based on quality-price evaluation only (ignoring all bias channels), joint exploitation
produces essentially no effect (-0.5\% vs +37.1\%) and complementarity
nearly vanishes (0.7 pp vs 19.7 pp). Documented AI
biases are therefore necessary for the exploitation mechanism: vertical tacit
collusion requires an exploitable target.

\vspace{1.5ex}

\textbf{Panel L: Heterogeneous Agent Populations (N=100 trials)}

\begin{longtable}[]{lllll}
\toprule
Population Mix & Joint Effect & Complementarity & Cohen's d & P-value \\
\midrule
\endhead
\bottomrule
\endlastfoot
100\% high-bias & +36.2\% & +18.4 pp & 2.18 & \textless0.0001 \\
75\% high / 25\% low & +35.0\% & +19.7 pp & 2.62 & \textless0.0001 \\
50\% high / 50\% low & +32.2\% & +18.9 pp & 2.32 & \textless0.0001 \\
25\% high / 75\% low & +28.1\% & +17.4 pp & 2.31 & \textless0.0001 \\
100\% low-bias & +21.9\% & +13.6 pp & 2.04 & \textless0.0001 \\
\end{longtable}

Complementarity persists across all population compositions. Even markets
dominated by low-bias consumers (75-100\%) show substantial super-additive
harm (d > 2.0), confirming the mechanism operates regardless of consumer
heterogeneity.

\vspace{1.5ex}

\textbf{Panel M: Position Bias Magnitude (Row Bonus) (N=50 trials)}

\begin{longtable}[]{lllll}
\toprule
Row Bonus & Joint Effect & Complementarity & Cohen's d & P-value \\
\midrule
\endhead
\bottomrule
\endlastfoot
0.30 (weak) & +24.8\% & +19.8 pp & 2.22 & \textless0.0001 \\
0.45 & +31.1\% & +23.3 pp & 3.06 & \textless0.0001 \\
0.60 & +32.7\% & +20.8 pp & 2.52 & \textless0.0001 \\
0.75 & +35.4\% & +21.0 pp & 2.74 & \textless0.0001 \\
0.90 (baseline) & +37.1\% & +19.7 pp & 2.75 & \textless0.0001 \\
\end{longtable}

Position bias magnitude determines how strongly the AI agent weights position
in rankings. Complementarity is robust across all magnitudes
(19.7--23.3 pp), with peak complementarity at moderate levels (0.45).

\vspace{1.5ex}

\textbf{Panel N: Quality Weight Sensitivity (N=50 trials)}

\begin{longtable}[]{lllll}
\toprule
Quality Wt & Joint Effect & Comp & Cohen's d & Notes \\
\midrule
\endhead
\bottomrule
\endlastfoot
0.10 & +19.7\% & -6.2 pp & -0.61 & Substitution \\
0.15 & +37.1\% & +19.7 pp & 2.75 & Baseline \\
0.25 & +27.5\% & +18.3 pp & 2.42 & Strong comp. \\
0.40 & +17.7\% & +12.0 pp & 2.03 & Moderate comp. \\
0.60 & +10.6\% & +6.9 pp & 1.63 & Reduced comp. \\
\end{longtable}

Quality weight determines how much the AI agent values product quality
versus other factors. At very low quality weight (0.10), complementarity
becomes negative: either actor alone can fully exploit. As quality
weight increases, harm decreases but complementarity persists until
agents sufficiently prioritize quality.

\vspace{1.5ex}

\textbf{Panel O: Platform Take Rate (N=50 trials)}

\begin{longtable}[]{lllll}
\toprule
Take Rate & Joint Effect & Complementarity & Cohen's d & P-value \\
\midrule
\endhead
\bottomrule
\endlastfoot
0\% & +37.1\% & +19.7 pp & 2.75 & \textless0.0001 \\
5\% & +22.2\% & +25.2 pp & 1.32 & \textless0.0001 \\
10\% & +14.3\% & +16.7 pp & 0.87 & \textless0.0001 \\
15\% & +10.0\% & +12.3 pp & 0.59 & \textless0.0001 \\
20\% & +10.0\% & +12.1 pp & 0.67 & \textless0.0001 \\
\end{longtable}

Platform take rate is the fraction of transaction value extracted by the
platform. At 0\% take rate (baseline), all seller revenue goes to
sellers. Introducing take rates reduces joint harm but complementarity
peaks at 5\% (+25.2 pp), suggesting moderate platform extraction
intensifies the collusive dynamic before seller margins become too thin.

\vspace{1.5ex}

\textbf{Panel P: Naive Seller Baselines (N=50 trials)}

\begin{longtable}[]{lllll}
\toprule
Seller Strategy & Joint Effect & Comp & Cohen's d & Notes \\
\midrule
\endhead
\bottomrule
\endlastfoot
Learning (baseline) & +37.1\% & +19.7 pp & 2.75 & Adaptive \\
Fixed bid=1 & +37.4\% & +34.9 pp & 1.39 & Amplifies \\
Fixed bid=2 & +37.4\% & +28.4 pp & 0.99 & Moderate \\
Fixed bid=1, manip=1 & +37.4\% & +31.0 pp & 0.89 & Combined \\
Fixed bid=1, manip=2 & +37.4\% & +36.8 pp & 1.13 & Maximum \\
\end{longtable}

When sellers use naive fixed strategies instead of learning, complementarity
\emph{increases} substantially (up to +36.8 pp vs +19.7 pp baseline). This
suggests learning sellers partially \emph{mitigate} exploitation by competing
away margins. Naive sellers amplify the platform's gatekeeper power.

\vspace{1.5ex}

\textbf{Panel Q: Convergence Speed (N=100 trials, 30,000 rounds)}

\begin{longtable}[]{llll}
\toprule
Threshold & Mean Rounds & Median Rounds & Trials Reaching \\
\midrule
\endhead
\bottomrule
\endlastfoot
5 pp complementarity & 2,070 & 2,000 & 100/100 \\
10 pp complementarity & 2,610 & 2,000 & 100/100 \\
15 pp complementarity & 3,430 & 2,000 & 100/100 \\
\end{longtable}

Vertical tacit collusion emerges rapidly. All 100 trials reach 15 pp
complementarity within 3,430 rounds on average (median: 2,000). The
exploitative equilibrium is not a slow-developing phenomenon: strategic
complementarity emerges within the first 10\% of simulation rounds. The complementarity trajectory over time is shown below:

\begin{longtable}[]{lll}
\toprule
Rounds & Complementarity & 95\% CI \\
\midrule
\endhead
\bottomrule
\endlastfoot
1,000 & +5.8 pp & [2.5, 9.2] \\
2,000 & +13.4 pp & [10.1, 16.7] \\
5,000 & +14.2 pp & [11.5, 17.0] \\
10,000 & +14.0 pp & [10.8, 17.2] \\
20,000 & +21.1 pp & [17.9, 24.3] \\
30,000 & +17.8 pp & [16.9, 18.6] \\
\end{longtable}

\vspace{1.5ex}

\textbf{Panel R: Gatekeeper Threshold Analysis (N=100 trials)}

\begin{longtable}[]{lllll}
\toprule
Bid Weight & Seller Effect & 95\% CI & Harm Rate & Cohen's d \\
\midrule
\endhead
\bottomrule
\endlastfoot
0.00 & $-$9.0\% & [$-$9.8, $-$8.1] & 4/100 & $-$2.10 \\
0.33 & $-$8.2\% & [$-$9.1, $-$7.3] & 7/100 & $-$1.77 \\
0.50 & $-$7.1\% & [$-$8.0, $-$6.2] & 10/100 & $-$1.55 \\
0.67 & $-$7.6\% & [$-$8.4, $-$6.8] & 4/100 & $-$1.86 \\
1.00 & +69.2\% & [+68.8, +69.6] & 100/100 & +34.90 \\
\end{longtable}

This analysis fixes platform bid weight while allowing sellers to learn,
testing the gatekeeper mechanism. Seller manipulation \emph{helps} consumers
(negative effect) at all bid weights up to 0.67. Only when the platform
completely abandons quality-based ranking (bid weight = 1.0) do seller
strategies flip to massively harmful (+69.2\%). The threshold is binary,
not gradual: any quality signal in rankings transforms seller behavior
from harmful to beneficial.

\vspace{1.5ex}

\textbf{Panel S: Equal Bias Weights Analysis (N=100 trials)}

\begin{longtable}[]{llllll}
\toprule
Config. & Pos. Bias & Comp & 95\% CI & $d$ & $P$ \\
\midrule
\endhead
\bottomrule
\endlastfoot
Baseline & 1.45 & +19.7 pp & [18.3, 21.1] & 2.75 & \textless0.0001 \\
Equal moderate (0.30) & 0.45 & +15.6 pp & [14.4, 16.8] & 2.53 & \textless0.0001 \\
Equal strong (0.50) & 0.73 & +24.4 pp & [22.6, 26.2] & 2.64 & \textless0.0001 \\
Position reduced & 1.00 & +21.4 pp & [19.7, 23.1] & 2.56 & \textless0.0001 \\
Quality boosted (3$\times$) & 1.45 & +10.6 pp & [9.5, 11.7] & 1.96 & \textless0.0001 \\
\end{longtable}

\textbf{Table S3. Comprehensive robustness analyses.} This analysis tests whether complementarity requires position bias to dominate
quality-price effects. Complementarity persists across all configurations:
when biases are equalized (15.6 pp), when position is reduced (21.4 pp), and
even when quality weight is tripled (10.6 pp). The mechanism reflects genuine strategic complementarity robust to
bias specification. See main text and Supplementary Notes 6--21 for detailed interpretation of each panel.

\printbibliography